\documentclass[twocolumn,twocolappendix]{aastex63}

\usepackage{lineno}
\usepackage{acronym}[acronymlists={hidden}]

\usepackage{amsmath}

\usepackage{xspace}

\usepackage{color}

\newcommand{\MESA}{{\texttt{MESA}}\xspace{}}
\newcommand{\COSMIC}{{\texttt{COSMIC}}\xspace{}}
\newcommand{\BSE}{{\texttt{BSE}}\xspace{}}

\newcommand{\qcrit}{$q_{\mathrm{crit}}$}

\newcommand{\fforward}{$f_{\mathrm{forward}}$}
\newcommand{\fbackward}{$f_{\mathrm{backward}}$}
\newcommand{\caseAHMXB}{Case-A~HMXB}
\newcommand{\caseAHMXBs}{Case-A~HMXBs}

\defcitealias{belczynski_compact_2008}{B+2008}

\newcommand{\be}{\begin{enumerate}}
\newcommand{\ee}{\end{enumerate}}

\shorttitle{}
\shortauthors{Gallegos-Garcia et al.}

\begin{document}

\title{Do High-spin High-mass X-ray binaries contribute to the population of merging binary black holes?}

\author[0000-0003-0648-2402]{Monica Gallegos-Garcia}
\affiliation{Department of Physics and Astronomy, Northwestern University, 2145 Sheridan Road, Evanston, IL 60208, USA}
\affiliation{Center for Interdisciplinary Exploration and Research in Astrophysics (CIERA),1800 Sherman, Evanston, IL 60201, USA}

\author[0000-0002-1980-5293]{Maya Fishbach}
\affiliation{Center for Interdisciplinary Exploration and Research in Astrophysics (CIERA),1800 Sherman, Evanston, IL 60201, USA}

\author[0000-0001-9236-5469]{Vicky Kalogera}
\affiliation{Department of Physics and Astronomy, Northwestern University, 2145 Sheridan Road, Evanston, IL 60208, USA}
\affiliation{Center for Interdisciplinary Exploration and Research in Astrophysics (CIERA),1800 Sherman, Evanston, IL 60201, USA}

\author[0000-0003-3870-7215]{Christopher P L Berry}
\affiliation{SUPA, School of Physics and Astronomy, University of Glasgow, Glasgow G12 8QQ, UK}
\affiliation{Center for Interdisciplinary Exploration and Research in Astrophysics (CIERA),1800 Sherman, Evanston, IL 60201, USA}

\author[0000-0002-2077-4914]{Zoheyr Doctor}
\affiliation{Center for Interdisciplinary Exploration and Research in Astrophysics (CIERA),1800 Sherman, Evanston, IL 60201, USA}

\begin{abstract}

Gravitational-wave observations of binary black hole (BBH) systems point to black hole spin magnitudes being relatively low.
These measurements appear in tension with high spin measurements for high-mass X-ray binaries (HMXBs).
We use grids of \MESA{} simulations combined with the rapid population-synthesis code \COSMIC{} to examine the origin of these two binary populations.
It has been suggested that Case-A mass transfer while both stars are on the main sequence can form high-spin BHs in HMXBs.
Assuming this formation channel, we show that depending on critical mass ratios for the stability of mass transfer, $48$--$100\%$ of these Case-A HMXBs merge during the common-envelope phase and up to $42\%$ result in binaries too wide to merge within a Hubble time.
Both \MESA{} and \COSMIC{} show that high-spin HMXBs formed through Case-A mass transfer can only form merging BBHs within a small parameter space where mass transfer can lead to enough orbital shrinkage to merge within a Hubble time. 
We find that only up to 11\% of these Case-A HMXBs result in BBH mergers, and at most $20\%$ of BBH mergers came from Case-A HMXBs. 
Therefore, it is not surprising that these two spin distributions are observed to be different.
\end{abstract}


\section{Introduction}

Correct interpretation of gravitational-wave (GW) data and a complete understanding of black hole (BH) spin predictions from stellar and binary evolution are crucial to reveal the formation channels of merging binary BHs (BBHs).
Of the BBH mergers detected by the LIGO Scientific, Virgo, and KAGRA Collaboration, most appear to have a small effective inspiral spin, $\chi_{\mathrm{eff}}\lesssim 0.2$--$0.3$ \citep{2021LIGO_GWTC-2.1,2021LIGO_GWTC-3}.
The effective inspiral spin is a mass-weighted combination of the spin components aligned with the orbital angular momentum \citep{Santamaria2010,Ajith2011}, and hence it can be difficult to disentangle the component BH spin magnitudes from the spin--orbit alignment.
Nevertheless, combining all the BBH mergers observed so far and fitting for the spin magnitude and tilt distributions, \citet{2021LIGO_population} found that component spin magnitudes tend to be smaller than $\chi_i \sim 0.4$, a feature that could have implications for the understanding BH natal spins. 
Other important but contended features of the BBH spin distribution include the possibility of a zero-spin excess \citep{Roulet2021,Galaudage2021}, and the presence of systems with spin--orbit misalignments larger than $90^\circ$  \citep[implying $\chi_{\mathrm{eff}}< 0$ ;][]{2021LIGO_population,LIGO2021_population_properties}. 
Implementing a series of hierarchical analyses of the BBH population, \citet{Callister2022} found preference for significant spin--orbit misalignment among the merging BBH population, but show that there is no evidence that GW data includes an excess of zero-spin systems. 
This latter point is in agreement with other studies \citep{Kimball2020,Kimball2021,Mould2022}, and indicates that the majority of merging BBHs have small but non-zero spins \citep{2021LIGO_population}.

The natal spins of BHs are largely determined by angular momentum (AM) transport from the core of the progenitor star to its envelope.
If this AM transport is assumed to be efficient, it acts to decrease the rotation rate of the core as the envelope expands and loses AM through winds, resulting in BHs born from single stars with spins of $\sim 10^{-2}$ \citep{Spruit1999,Fuller2015,FullerMa2019}.
Evidence for efficient AM transport comes, in part, from comparison to observations of neutron star and white dwarf spins \citep{Heger2005,Suijs2008}.
However, we currently lack unambiguous evidence that AM transport is efficient in more massive stars, especially since there is no observed excess of zero-spin systems in GW data.
Additionally, \cite{Cantiello2014} found that this mechanism fails to reproduce the slow rotation rates of the cores of low-mass stars, which led to a revision of the AM transport process \citep{Fuller2019_slowing_spins}. 
To further complicate this story, failed SN explosions can alter the spin of a new-born BH \citep{Batta2017,Schroder2018,Batta2019}, and binary evolution after the first BH is formed, like tidal synchronization, can increase the spin of the second-born BH, provided that the orbit is tight enough \citep{Qin2018,Bavera2020_origin_spin_BBHs,Fuller2022}. 

High-mass X-ray binaries (HMXBs) consist of a compact object, either a neutron star or BH, with a massive donor star $\gtrsim 5 M_{\odot}$ \citep{Remillard2006,vandenHeuvel2019}.
Our focus is on HMXBs with BH accretors, and we refer to these as \emph{HMXBs} henceforth.
Of the three HMXBs with confident BH spin measurements (M33 X-7, Cygnus X-1 and LMC X-1), all BHs are observed to have high spins, with spin magnitudes $\chi \gtrsim0.8$ \citep{Liu2008,Miller-Jones2021,Reynolds2021}.
Although there are only three of these systems, it is clear that they have a distinct spin distribution compared to merging BBHs~\citep{RouletZaldarriaga2019,Reynolds2021,Fishbach2022}.

We might naively expect that for both HMXBs and merging BBH systems, the spin of the first-born BH represents its natal spin.
As discussed above, BH spins can be altered during a SN event or by strong binary interactions such as tides, which are likely to be more important for the second-born BH.
While BBHs can be expected go through a HMXB phase, not all HMXBs will evolve to form merging BBHs  \citep[e.g.,][]{Belczynski2011,Belczynski2012,Miller-Jones2021,Neijssel2021_cygX1}.
One goal of our study to find an evolutionary path that can explain current observations: one that can impart large spin on the first-born BH in HMXBs but not in merging BBHs.

We must consider the possibility that these two classes of binaries may only appear different due to the limitations of how they are observed.
\citet{Fishbach2022} investigated whether the differences in the mass and spin distributions of HMXBs and merging BBHs may be a result of GW observational selection effects alone.
Based upon GWTC-2 observations \citep{Abbott2020}, they found that, accounting for GW observational selection effects and the small-number statistics of the observed HMXBs, the masses of the observed HMXBs are consistent with the BBH mass distribution.
However, considering BH spins, the merging population of BBHs may include only a small subpopulation of systems that are HMXB-like (systems containing a rapidly spinning component with $\chi \gtrsim 0.8$, and preferentially aligned with the orbital angular momentum axis, as expected from isolated binary evolution). 
Conservatively, \citet{Fishbach2022} find that a HMXB-like population can make up at most $30\%$ of merging BBH systems.
It is therefore important to understand how the specific evolutionary pathways of merging BBHs and HMXBs shape their observed spins distributions \citep{Liotine2022}.

We investigate if high-spin HMXBs are expected to contribute to the population of merging BBHs by modeling the evolution of these binaries.
Henceforth we refer to the population of BBH systems that merge within a Hubble time as \emph{BBHs}, except, in cases where it can lead to confusion, we use \emph{merging BBHs} for clarity.
To identify high-spin HMXBs in simulations, we assume the spin of the first-born BH is imparted by the scenario of Case-A mass transfer (MT) while both stars are on the main sequence \citep[MS;][]{Valsecchi2010Nature,Qin2019}.
In this scenario, the donor star, which is also the progenitor of the first-born BH, could form a high-spin BH following a combination of ({\romannumeral 1}) MT that prevents significant radial expansion; ({\romannumeral 2}) strong tidal synchronization at low orbital periods, and ({\romannumeral 3}) inefficient AM transport within the massive star post MS.
We do not follow the spin evolution of these BH progenitors, but simply assume that systems following this Case-A MT formation path can form a (near) maximally spinning first-born BH \citep{Qin2019}. 
We refer to these high-spin HMXBs as \emph{\caseAHMXBs{}}.
We show that only a minority of  \caseAHMXBs{} result in BBHs. 
Similarly, only a small fraction of BBHs had a \caseAHMXB{} progenitor. 
This implies that the BHs observed in HMXBs and those in BBHs predominantly belong to different astrophysical populations. 

This work is organized as follows. 
In Section~\ref{sec:methods} we outline our procedure for combining \MESA{} and \COSMIC{} simulations and provide an overview of the stellar and binary physics parameters used. 
In Section~\ref{sec:results} we quantify how many \caseAHMXBs{} form BBHs, and what fraction of our total BBHs in the population had \caseAHMXB{} progenitors (Appendix~\ref{sec:additional_models} includes results for additional models). 
In Section~\ref{sec:discussion} we discuss caveats and avenues for future work. 
We summarize our findings in Section~\ref{sec:conclusion}. 
In Appendix~\ref{sec:alternative_formation} we review a few alternative channels for forming a high-spin BH as the first
born BH in the binary and their possible contributions to the merging BBH population.

\section{Method}\label{sec:methods}

We combine detailed binary evolution simulations modeled using \MESA{} \citep{Paxton2011,Paxton2013,Paxton2015,Paxton2019} with simulations using the rapid population-synthesis code \COSMIC{} \citep{Breivik2020}, which is based upon the evolutionary models of \BSE{} \citep{Hurley2002}, to determine if Case-A HMXBs and BBHs originate from distinct populations.
This combination allows us to simulate large populations of binaries, and assess whether our results are robust by comparing them to populations informed by detailed simulations. 
Our simulations are computed using version 12115 of \MESA{}, and version 3.4 of \COSMIC{}.
Our procedure for combing \COSMIC{} and \MESA{} simulations is similar to \citet{Gallegos-Garcia2021}.
Here we provide a brief summary and highlight any minor differences. 

\begin{figure}[!htb]
\centering
\includegraphics[width=0.45\textwidth]{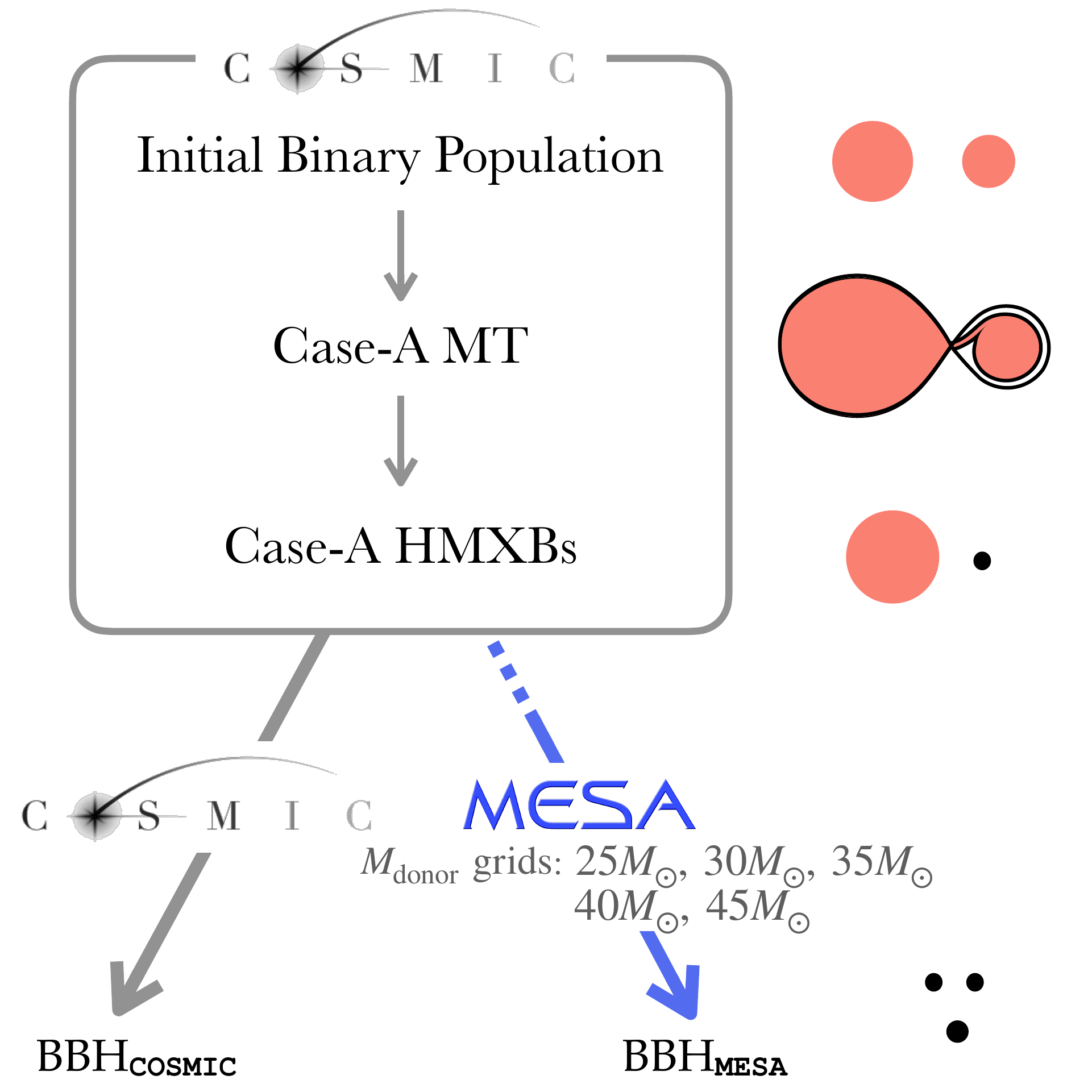}
\caption{Illustration of method.
The evolution of all binaries, from an initial ZAMS population, through Case-A MT while both stars are on the MS, to the formation of \caseAHMXBs{}, is simulated entirely with \COSMIC{}. 
Starting from this population of \caseAHMXBs{}, we match each \caseAHMXBs{} to the nearest binary simulation in terms of orbital period and mass ratio from our grids of \MESA{} simulations.
For comparison, we use both \COSMIC{} to simulate the remaining evolution.
} 
\label{fig:method}
\end{figure}

We generate an initial population of binaries with \COSMIC{} with multidimensional initial binary parameters following \cite{MoeDiStefano2017}. 
We evolve these binaries from zero-age MS (ZAMS) until the formation of a hydrogen-rich donor with a BH companion (BH--H-rich star). 
We refer to this as the \emph{HMXB} stage. 
We do not explicitly consider the criteria for the formation
of an accretion disc or the observability of the X-ray flux \citep[e.g.,][]{Hirai2021}.
In this population, we highlight the systems that undergo Case-A MT while both stars are on the MS because these may result in high-spin HMXBs \citep[\caseAHMXBs{};][]{Valsecchi2010Nature,Qin2019}.
To compare our results across  different donor masses at the BH--H-rich star stage, we separate these binaries into subpopulations determined by the donor mass.
We consider five mass ranges in our \COSMIC{} simulations, $M_{\mathrm{donor}} =$ $(25 \pm 2.5) M_{\odot}$, $(30 \pm 2.5) M_{\odot}$, $(35 \pm 2.5) M_{\odot}$, $(40 \pm 2.5) M_{\odot}$, and $(45 \pm 2.5) M_{\odot}$.
We use a grid of \MESA{} simulations at a \emph{single} donor mass to compare to a selected mass range of \COSMIC{} systems: i.e., a mass range of $M_{\rm donor} = (35 \pm 2.5)~M_{\odot}$ in our \COSMIC{} models is compared to a single grid of \MESA{} simulations with $M_{\rm donor} = 35 M_{\odot}$. 
We also approximate all H-rich stars in \COSMIC{} as MS stars in our \MESA{} simulations.
To determine which systems form BBHs, the HMXB population is then evolved to end of life with both \COSMIC{} and with nearest neighbor interpolation in terms of orbital period and mass ratio of the \MESA{} runs following \citet{Gallegos-Garcia2021}.
A schematic of our method is shown in Figure~\ref{fig:method}.

For each subpopulation, we label different final outcomes for \caseAHMXBs{}, which includes those that form BBHs.
From this we calculate \fforward{}, the fraction of systems that result in each of the outcomes. 
We also calculate \fbackward{}, the fraction of BBHs that had a \caseAHMXB{} progenitor and are thus candidates for BBHs with at least one high-spin BH. 

\subsection{Stellar \& Binary Physics} \label{subsection:star physics}

We make use of the grids of \MESA{} simulations from \cite{Gallegos-Garcia2021}, and calculate an additional grid of simulations with $M_{\mathrm{donor}}=45 M_{\odot}$. 
Our models are initialized at a metallicity $Z = 0.1 Z_{\odot}$, defining $Z_{\odot} = 0.0142$ and $Y_{\odot}=0.2703$ \citep{2009Asplund}. 
We also simulate one model at solar metallicity.
We specify the helium fraction as $Y = Y_\mathrm{Big\ Bang} +  ( Y_{\odot} - Y_\mathrm{Big\ Bang} ) Z/Z_{\odot}$, where $Y_\mathrm{Big\ Bang} = 0.249$ \citep{Planck2016}. 
For simulations run with \COSMIC{}, the stellar and binary physics parameters are the same as in \citet{Gallegos-Garcia2021}, except now all simulations are updated to have MT prescriptions from \citet{Claeys2014}.

As in \citet{Gallegos-Garcia2021}, we carefully maintain consistency among the stellar and binary physics parameters between the two codes.
The \COSMIC{} wind prescription most similar to the prescription used in our \MESA{} simulations treats O and B stars following \citet{Vink2001_hot_highH}, and Wolf--Rayet stars following \citet{Hamann1998} reduced by factor of $10$ \citep{Yoon2010} with metallicity scaling of $(Z/Z_{\odot})^
{0.86}$ \citep{Vink2005}. 
For the formation of BHs, when \MESA{} models reach core carbon depletion (central $^{12}$C abundance $< 10^{-2}$), they are assumed to undergo direct core collapse to a BH with mass equal to their baryonic mass.
In \COSMIC{}, we follow the Delayed prescription of \citet{Fryer2012}. 
We expect the small differences between the winds and supernova prescriptions for \MESA{} and \COSMIC{} to not significantly affect results.

Our method for identifying high-spin HMXBs relies on Case-A MT while both stars are still on the MS.
In \citet{Qin2019}, this scenario was modeled using detailed \MESA{} simulations that focused on the MT episode and binary evolution before the first BH was formed.
In our study, we only model this Case-A MT stage of evolution with \COSMIC{}, which likely results in differences between simulations performed with \MESA{}.
In a preliminary study, over a small parameter space in donor mass and orbital period, we found that in some cases, simulations ran with \COSMIC{} tended to overestimate the number of \caseAHMXBs{} by roughly an factor of two compared to Figure~2 in \citet{Qin2019}.
We therefore treat the \caseAHMXBs{} populations in \COSMIC{} as upper limits.

The evolution of Case-A MT occurs at low initial orbital periods ($\lesssim 25~\mathrm{days}$).
At these periods, common envelope (CE) evolution is expected to be unsuccessful at removing the envelope given the energy budget formalism \citep{vandenHeuvel1976,Webbink1984, Ivanova2013}.
As a result of this, BBH mergers can only form through stable MT, or chemically homogeneous evolution \citep[CHE;][]{Marchant2016,deMink2016}.
The mass-ratio threshold \qcrit{} that sets the stability of MT for these donors (i.e., whether a system undergoes CE) therefore determines how many systems will be able to form BBHs through stable MT.
If the mass ratio $q=M_{\mathrm{accretor}}/M_{\mathrm{donor}}$ is less than \qcrit{}, the system enters unstable MT and a CE forms. 
A smaller \qcrit{} value means fewer systems undergo CE.
To explore uncertainties in this part of binary evolution, in the \COSMIC{} models presented here, we vary the critical mass ratios by considering three different $q_{\rm crit}$ prescriptions following \citet{belczynski_compact_2008}, \citet{neijssel_effect_2019}, and \citet{Claeys2014}: the \citet{belczynski_compact_2008} prescriptions are used for the results shown in Section~\ref{sec:results} while the other results are shown in Appendix~\ref{sec:additional_models}. 
This choice of critical mass ration is separate from the MT prescription, which sets the rate of mass lost from the donor star and follows \citet{Claeys2014} for all \COSMIC{} simulations.  

Case-A MT between two MS stars is the first evolutionary phase where \qcrit{} becomes important in our simulations. 
We denote this first critical mass ratio as $q_{{\mathrm {crit}}}^{\mathrm{MS}}$.
Out of the set of prescriptions we consider, the model following \citet{belczynski_compact_2008} allows more MS stars to proceed with stable MT instead of CE.
For this model, all H-rich donors in binaries with $q$ larger than $q_{\mathrm{crit}}^{\mathrm{MS}} = 0.33$ are assumed to be stable. 
\citet{neijssel_effect_2019} has the second largest value with $q_{\mathrm{crit}}^{\mathrm{MS}}=0.58$.
This is followed by \cite{Claeys2014}, which uses $q_{\mathrm{crit}}^{\mathrm{MS}}=0.625$.
The differences among $q_{\mathrm{crit}}^{\mathrm{MS}}$ are important, as they can affect the resulting population of \caseAHMXBs{}.

Equally as important are the $q_{\rm crit}$ values for Roche lobe overflow onto a BH during the HMXB phase, which we denote as $q_{\mathrm {crit}}^{\mathrm{BH}}$. 
Generally, H-rich stars include Hertzsprung gap (HG), first giant branch, core helium burning, early asymptotic giant branch (AGB), and thermally pulsing AGB stars, but for the population of \caseAHMXBs{}, the most evolved H-rich star in our BH--H-rich star population is a HG star.
For systems containing BH--HG stars, the \citet{Claeys2014}, \citet{neijssel_effect_2019} and \citet{belczynski_compact_2008} prescriptions use $q_{\mathrm{crit}}^{\mathrm{BH}}=0.21$, $q_{\mathrm{crit}}^{\mathrm{BH}}=0.26$ and $q_{\mathrm{crit}}^{\mathrm{BH}}=0.33$, respectively.
Values similar to the last were also derived by \cite{Tauris2000}, \cite{Hurley2000}, and \cite{Pavlovskii2017}.

\section{Results} \label{sec:results}

Here we show the outcomes of \caseAHMXBs{}, i.e., binaries that are assumed to be candidates for high-spin HMXBs following a phase of Case-A MT while both stars are on the MS (Section~\ref{subsection:outcomes}).
We also quantify how many of these \caseAHMXBs{} form BBHs, and what fraction of the total BBHs in the population had \caseAHMXB{} progenitors (Section~\ref{subsection:fractions}). 

\subsection{Outcomes of \caseAHMXBs{}}
\label{subsection:outcomes}

\begin{figure*}[!htb]
\centering
\includegraphics[width=0.9\textwidth]{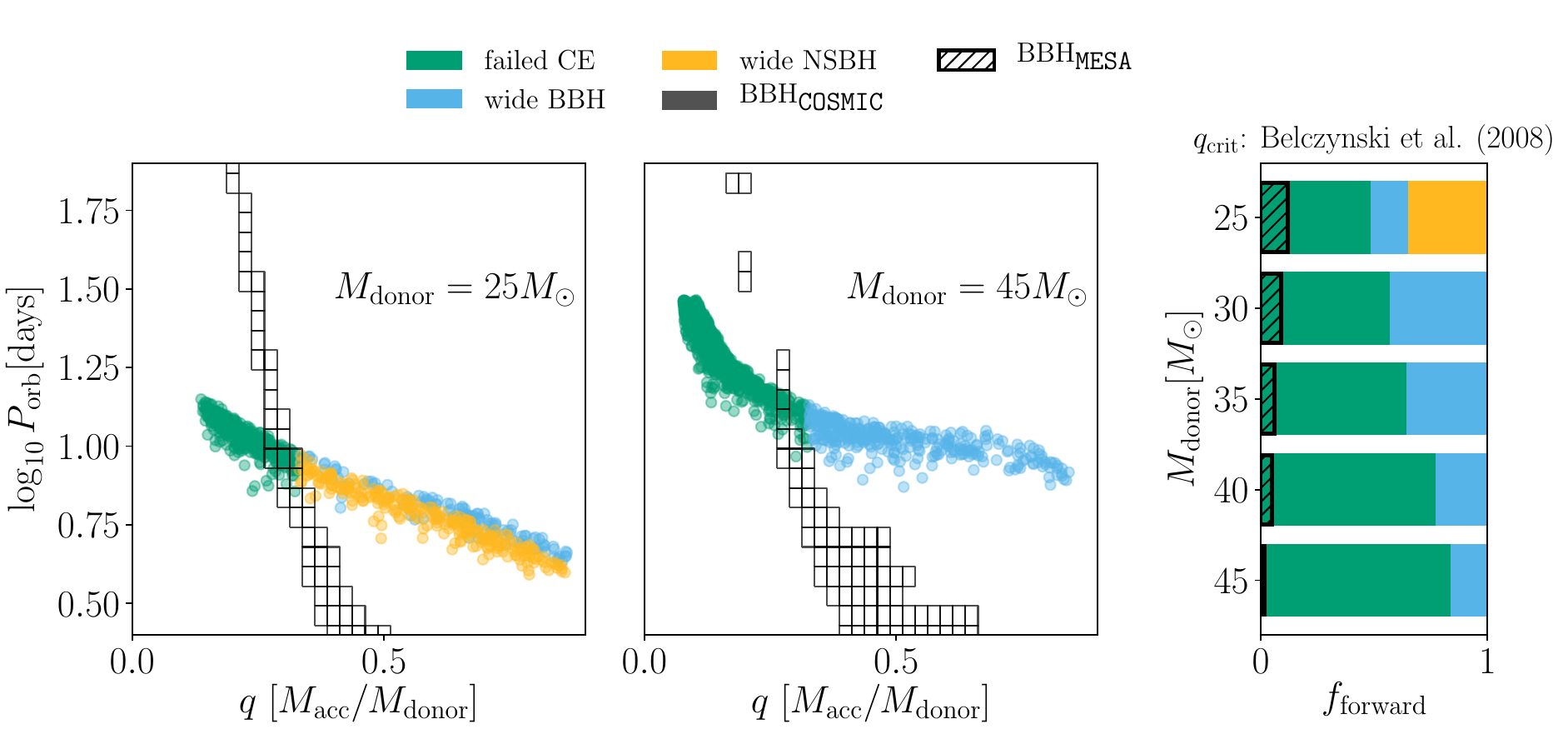}
\caption{Summary of outcomes for model with $q_{\mathrm{crit}}$ following \cite{belczynski_compact_2008} at $Z_{\odot}/10$. 
Points correspond to simulation outcomes for binaries ran with \COSMIC{}. 
The left panel corresponds to donor masses within the range $M_{\rm donor} = (25 \pm 2.5) M_{\odot}$, and the middle panel corresponds to $M_{\rm donor} = (45 \pm 2.5) M_{\odot}$. 
In these panels, black rectangles correspond to the parameter space where the corresponding grid of \MESA{} simulations for that donor mass result in BBHs. 
The right panel shows the fractions of each outcome as a function of donor mass. 
The hatched black bar corresponds to the fraction of BBHs for each donor mass given the grids of simulations ran with \MESA{}. 
In all three panels, binaries that merged during CE are shown in green, systems that resulted in wide NSBHs are in yellow, and wide BBHs are in light blue. } 
\label{fig:startrack_claeys_fractions}
\end{figure*}

We label four different final outcomes for \caseAHMXBs{} for models simulated with \COSMIC{}, and one outcome for the grids of \MESA{} simulations. 
These outcomes are the following. 
\begin{enumerate}
    \item Binaries that merge during CE. 
    These binaries are concentrated at unequal mass ratios $q$ for all masses and model variations. 
    We label them as \emph{failed CE}.
    \item Binaries that result in wide neutron star--BHs (NSBHs) that will not merge within a Hubble time.  
    This outcome only occurs for the least massive donor and we label them as \emph{wide NSBHs}. 
    \item \emph{Wide BBHs} that will not merge within a Hubble time.
    These systems make up most of the remainder of the binaries that do not merge during CE.
    \item Binaries that result in BBHs that merge within a Hubble time. 
    We label them as \emph{BBH$_{\COSMIC{}}$}. 
    \item We label \COSMIC{} \caseAHMXBs{} that result in BBHs following the nearest neighbors matching with the grids of \MESA{} simulations as \emph{BBH$_{\MESA{}}$}.
\end{enumerate}
The comparison between BBH$_{\COSMIC{}}$ and BBH$_{\MESA{}}$ allows us to both asses how detailed models of binary evolution affect the final outcome of \caseAHMXBs{} and test the robustest of our final results. 

Figure~\ref{fig:startrack_claeys_fractions} shows the final outcomes following $q_{\mathrm{crit}}$ prescriptions by \citet{belczynski_compact_2008}.
We show systems with H-rich donor masses within the range $M_{\mathrm{donor}} = (25 \pm 2.5)M_{\odot}$ and $(45 \pm 2.5)M_{\odot}$ on the left and middle panels, respectively. 
Each point in Figure \ref{fig:startrack_claeys_fractions} corresponds to a binary simulated with \COSMIC{}, with the color representing the final outcome as described above.  
The outcomes are plotted as a function of mass ratio $q$ and orbital period $P_{\mathrm{orb}}$ when the system became a BH--H-rich star, which is the starting state of the \MESA{} simulations.
On these same panels, the black rectangles show where our grids of BH–-MS \MESA{}  models result in BBHs.
In the right panel of Figure~\ref{fig:startrack_claeys_fractions} we also show the fractions of the final outcomes \fforward{} as a function of donor mass.
The hatched bars in this panel correspond to BBH$_{\MESA{}}$, the fraction of BBHs assumed to form after combining our grids \MESA{} simulations with the \COSMIC{} \caseAHMXB{} population. 
The binaries that make up this fraction are those that fall within the black rectangles. 
For this model, when simulating binary evolution entirely with \COSMIC{} we do not find any BBHs: BBH$_{\COSMIC{}}=0$. 
When combining \MESA{} with \COSMIC{} simulations we find that only a small fraction, at most $\sim12$\%, result in BBHs.
When considering all systems in this model, $M_{\mathrm{donor}} = (25\pm 2.5)$--$(45\pm 2.5)M_{\odot}$, only 5\% of binaries result in BBHs. 
The differences in BBH$_{\COSMIC{}}$ and BBH$_{\MESA{}}$ for this model are because some \caseAHMXBs{} that undergo failed CE with \COSMIC{} go through stable MT according to our grids of \MESA{} simulations.
In Appendix \ref{sec:additional_models} we present similar calculations for models using \qcrit{} following \citet{neijssel_effect_2019} and \citet{Claeys2014}: we find the similar values for BBH$_{\COSMIC{}}$ and BBH$_{\MESA{}}$ with these models (Table~\ref{table:HMXB_outcomes}).

In addition to varying \qcrit{}, we also simulated a population of binaries at solar metallicity and found no BBHs with \caseAHMXBs{} progenitors with either \COSMIC{} or \MESA{}. 
This is likely due to stronger winds at solar metallicities implemented in both codes that widen the orbits and reduce the number of BBHs.
We also assessed whether the fractions of \caseAHMXBs{} resulting in BBHs are affected by different initial binary parameter distributions.
Choosing each initial ZAMS parameter of the binary independently rather than choosing them jointly as in our default \cite{MoeDiStefano2017} initial distributions
we find a negligible change for BBH$_{\MESA{}}$ the model following \citet{belczynski_compact_2008}.

\newpage{}

\subsection{Fraction of high-spin BBHs}
\label{subsection:fractions}

Although we find that only a small fraction of \caseAHMXBs{} form BBHs, it is possible that this population of BBHs is large enough to contribute significantly to the full BBH population.
In addition to determining the fates of \caseAHMXBs{}, we must also consider the fraction of BBHs that had a \caseAHMXB{} progenitor, \fbackward{}.

For the model using \qcrit{} following \cite{belczynski_compact_2008}, we can only calculate \fbackward{} for binaries that we modeled with \MESA{} simulations, as BBH$_{\COSMIC{}}=0$.
We find \fbackward{} values between $0.05$--$0.2$, with the maximum value corresponding to donors with masses within the range $M_{\mathrm{donor}}=(45\pm 2.5)M_{\odot}$.
A summary of these values for the three \qcrit{} models is presented in Appendix~\ref{sec:fraction_highspin_BBH_appendix} (Table~\ref{table:f_backwards}). 
For all models, these fractions tend to be small ($<0.20$) which indicates that \caseAHMXB{} systems and BBHs likely have little association. 

\section{Discussion} \label{sec:discussion}

Here we discuss a few caveats in our study and a possible avenue for improvement. 
Further discussion of alternative formation scenarios for high-spin BHs is given in Appendix~\ref{sec:alternative_formation}. 

While we investigated whether different criteria for the stability of MT, $q_{\mathrm{crit}}$, affect our results (Appendix~\ref{sec:additional_models}), the set of prescriptions used are not exhaustive.  
Recent prescriptions, such as in \citet{olejak2021impact_CE_criteria}, were not examined.
Since the formation of Case A-HMXBs occurs over a small orbital period range, and our grids of \MESA{} simulations form BBHs over a small mass-ratio range at those orbital periods, the parameter space where Case A-HMXBs can lead to BBHs is small. 
Therefore, we do not expect significant differences in the fractions presented here with alternative \qcrit{} prescriptions.

For the modeling of binary evolution, we performed simulations of BH--H-rich star binaries with \MESA{}, but we simulated MS--MS evolution with \COSMIC{}.
Similar to comparing results of BH--H-rich star outcomes in \COSMIC{} to those from our \MESA{} simulations, it is important to also study the prior evolution of these binaries with detailed simulations. 
Our results may be affected by a better implementation of MT during MS--MS evolution, and when this MT becomes unstable leading to CE.

The modeling of MS--MS evolution with \COSMIC{} does not enable an adequate estimate of the star's core spin.
As a result, we did not follow the spin evolution of the BH progenitor in our simulations. 
With these limitations, we have only considered the Case-A MT (while both stars on the MS) scenario for forming high-spin HMXBs.
Since it is plausible that not all \caseAHMXBs{} will reach high-spin values, our results should be considered conservative upper limits. 
Additionally, we do not consider other spin-up mechanisms and their contributions.

Most of the shortcomings associated with the need for detailed simulations can be well-addressed with population-synthesis codes like \texttt{POSYDON} \citep{Fragos2022} that use \MESA{} simulations to model the full evolution of binary systems.
This would also allow future studies to include higher-mass progenitors than those considered here as they simulate binary evolution with ZAMS stars up to $120M_{\odot}$.

Finally, given the short orbital periods, it is plausible that \caseAHMXBs{} can not only form BBHs with one high-spin component, but perhaps impart non-negligible spin to the second-born BH through tides \citep{Qin2018,Bavera2020_origin_spin_BBHs}. 
A more detailed study concerning the spin evolution of the second-born BH from \caseAHMXBs{} may help constrain the observational features expected from this small population of BBHs in GW data.

\section{Conclusions}\label{sec:conclusion}

We have used grids of \MESA{} simulations combined with the rapid population-synthesis code \COSMIC{} to assess whether HMXBs with high-spin BHs and merging BBHs (referred to as BBHs) originate from distinct populations. 
To identify high-spin BHs in HMXBs, we adopted the scenario modeled in \citet{Qin2019}, which shows that Case-A MT while both stars are on the MS can result in a first-born BH that has high spin, as long as angular momentum transport in the star is inefficient. 
For BHs formed outside of this Case-A MT scenario, we assume that they will have distinctively lower spin than our \caseAHMXBs{}.

Our main conclusions are:
\begin{enumerate}
    \item  \caseAHMXBs{} do not tend to form BBHs.
    When using only \COSMIC{} simulations to model the full binary evolution, we find that at most 2\% of \caseAHMXBs{} result in BBHs. 
    When combining the \COSMIC{} population with grids of BH--H-rich star \MESA{} simulations, we find at most 12\% form BBHs.
    
    \item \caseAHMXBs{} contribute only a small fraction to the total merging BBH population.
    When considering all the merging BBHs for the range of masses investigated here, only 7\% had a \caseAHMXB{} progenitor.
    When considering the individual mass ranges, the most massive H-rich donor, $M_{\mathrm{donor}}= (45 \pm 2.5) M_{\odot}$, had the largest fraction with at most 20\% of BBHs having a \caseAHMXB{} progenitor.
    
    \item The scenario of Case-A MT while both stars are on the MS allows for the formation of high-spin HMXBs while forming a minority of BBHs, such that the expected population of GW sources would contain primarily low-spin BHs. 
\end{enumerate}
Although a fraction of \caseAHMXBs{} can result in BBHs, their formation path can be significantly different from the larger BBH population. 
These differences, which can lead to high-spin BHs, are important to consider when interpreting observations.

Our conclusions are in agreement with \citet{Fishbach2022}, who found that a subpopulation comprising of at most $30\%$ of BBHs may have features resembling rapidly spinning HMXB-like systems, where one BH component is high-spin. 
This is also in agreement with \cite{Neijssel2021_cygX1}, who, following a case study of Cygnus X-1 and finding a 5\% probability that it will result in a merging BBH within a Hubble time, infer that a small fraction of HMXBs like Cygnus X-1 may form BBHs. 

In our \COSMIC{} models we varied the mass ratio threshold for MT stability (Appendix \ref{sec:additional_models}) as this value determines which systems avoid CE and therefore lead to more Case-A MT systems and merging BBHs within a Hubble time.
We found that different MT stability prescriptions produce significantly different populations of \caseAHMXB{} systems. 
However, the $q_{\mathrm{crit}}$ prescriptions produce robust conclusions and can be consistent our grids of \MESA{} simulations.
Our results also remained similar when varying metallicity in one model and the initial ZAMS binary parameters.

Upcoming GW data will better resolve the spin distribution of BBHs, and as HMXB measurements improve we will have more accurate measurements of BH masses and spins in these systems.
With both types of observations constraining different aspects of binary evolution, combining information from both will provide a more complete understanding of the physics of binary evolution.
We can use studies like these to more accurately interpret these observed spins and to better understand the scenarios that lead to different stellar populations.

\begin{acknowledgements}

The authors thank Meng Sun for their feedback and assistance with our \MESA{} simulations and Katie Breivik for help with \COSMIC{}.
We thank Jeff Andrews, Michael Zevin, Ariadna Murguia Berthier, Aldo Batta, Katie Breivik and Will Farr for insightful conversations.
M.G.-G.\ is grateful for the support from the Ford Foundation Predoctoral Fellowship. 
M.F.\ is supported by NASA through NASA Hubble Fellowship grant HST-HF2-51455.001-A awarded by the Space Telescope Science Institute.
C.P.L.B.\ and Z.D.\ are grateful for support from the CIERA Board of Visitors Research Professorship.
V.K.\ is supported by a CIFAR G+EU Senior Fellowship, by the Gordon and Betty Moore Foundation through grant GBMF8477, and by Northwestern University.
This work utilized the computing resources at CIERA provided by the Quest high performance computing facility at Northwestern University, which is jointly supported by the Office of the Provost, the Office for Research, and Northwestern University Information Technology, and used computing resources at CIERA funded by NSF PHY-1726951.

Input files and data products are available for download from Zenodo.\footnote{\href{https://doi.org/10.5281/zenodo.6954573}{doi.org/10.5281/zenodo.6954573}}
\end{acknowledgements}

\software{ \MESA{} \citep{Paxton2011,Paxton2013,Paxton2015,Paxton2019};
\COSMIC{} \citep{Breivik2020};
\texttt{Matplotlib}  \citep{Hunter2007}; 
\texttt{NumPy}  \citep{vanderwalt2011};
\texttt{Pandas}  \citep{mckinney-proc-scipy-2010}.
}

\appendix

\section{Additional Models}\label{sec:additional_models}

In this Appendix we include the results using additional models. 
In comparison to the results using the \citet{belczynski_compact_2008} prescriptions for \qcrit{} shown in Section~\ref{sec:results}, here we discuss results using the \citet{neijssel_effect_2019} and \citet{Claeys2014} prescriptions.

\subsection{Outcomes of \caseAHMXBs{}}

\begin{figure*}[!htb]
\centering
\includegraphics[width=0.9\textwidth]{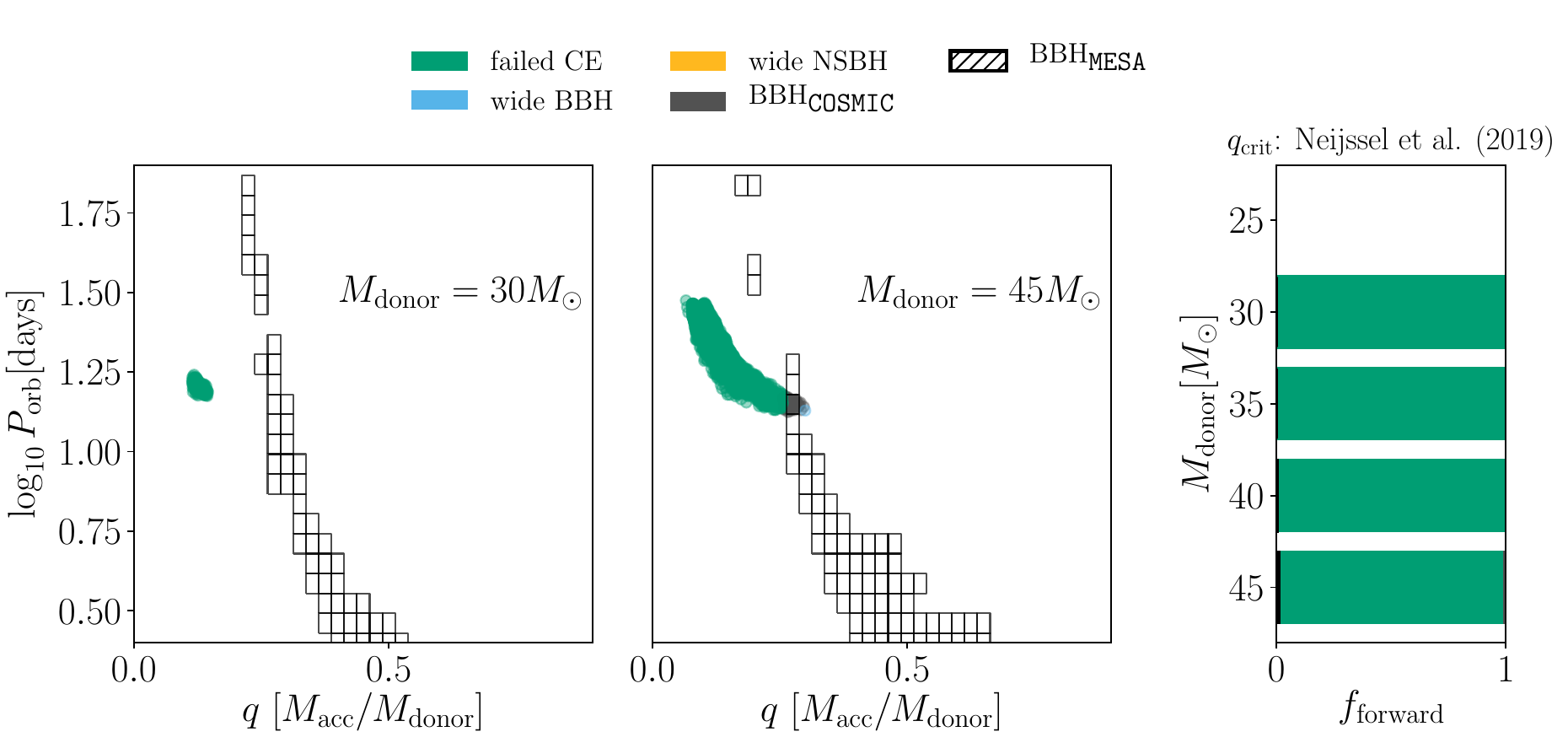}
\caption{Same as Figure~\ref{fig:startrack_claeys_fractions} but for the model using $q_{\mathrm{crit}}$ values following \citet{neijssel_effect_2019} and BBH mergers within a Hubble time are shown in gray. 
Binaries with donor masses in the range of $M_{\rm donor} = (30 \pm 2.5) M_{\odot}$ are shown in the left panel and $M_{\rm donor} = (45 \pm 2.5) M_{\odot}$ are shown in the middle panel. 
Although this model results in BBHs in the same parameter space as our grid of \MESA{} simulations, this outcome contributes only $0.01\%$ to the total outcome of \caseAHMXBs{}. 
} 
\label{fig:neijssel_claeys_fractions}
\end{figure*}

Figure~\ref{fig:neijssel_claeys_fractions} shows the same results as in Figure~\ref{fig:startrack_claeys_fractions} but for the model using $q_{\mathrm{ crit}}$ following \citet{neijssel_effect_2019}. 
We show binaries with donor masses within the range $M_{\rm donor}~=~(30 \pm 2.5) M_{\odot}$ and $M_{\rm donor}=(45 \pm 2.5) M_{\odot}$ on the left and middle panels respectively.
In this model, no \caseAHMXBs{} form within the mass range  $M_{\rm donor} = (25 \pm 2.5) M_{\odot}$.
This is likely due to the larger $q_{\mathrm{ crit}}^{\mathrm{MS}}$ value used in the first phase of MT. 
This larger value intrinsically limits binaries with less massive secondary stars, which would otherwise become the donors in the HMXB phase, from proceeding with stable MT during the first MT phase.
This model has a lower $q_{\mathrm{crit}}^{\mathrm{BH}}$ value compared to \citet{belczynski_compact_2008} and allows more BH--H-rich systems to proceed with stable MT when the donor is a HG star.
For donors with masses within the range $M_{\rm donor} = (45 \pm 2.5) M_{\odot}$, this results in BBHs following stable MT only (gray points in middle panel).
Additionally, at this donor mass, the BBHs modeled with \COSMIC{} are consistent with the parameter space where our \MESA{} simulations result in BBHs (the overlap of gray points and black rectangle).
This is a small region in parameter space for both \COSMIC{} and \MESA{} with a width in mass ratio $\Delta q \sim 0.05$ and $0.0625~\mathrm{dex}$ in orbital period.
Compared to Figure~\ref{fig:startrack_claeys_fractions}, the range of mass ratios of \caseAHMXBs{} is smaller, spanning $q \approx 0.1$--$0.3$ compared to $q \approx 0.1$--$0.8$. 
This smaller range in $q$ decreases the number of BBHs over all donor masses when the \COSMIC{} \caseAHMXB{} population is combined our grids of \MESA{} simulations.
This can be seen in the right-most panels of Figure~\ref{fig:startrack_claeys_fractions} and Figure~\ref{fig:neijssel_claeys_fractions}. 
Although the \COSMIC{} \caseAHMXB{} population is different for these two models, we find similar results for the fraction of \caseAHMXBs{} that result in BBHs.
As in the model using $q_{\mathrm{ crit}}$ following \citet{belczynski_compact_2008}, this model does not result in a significant fraction of BBHs.

In our third model we use $q_{\rm crit}$ prescriptions following \citet{Claeys2014}.
This model results in similar BBH factions and qualitatively similar \caseAHMXB{} populations to the model using $q_{\rm crit}$ following \citet{neijssel_effect_2019}.
The \caseAHMXB{} populations for this model have smaller mass ratio range with $q \approx 0.1$--$0.25$.
As a result, unlike the model using $q_{\rm crit}$ from \citet{neijssel_effect_2019}, we do not find an overlapping region between \COSMIC{} BBHs from the \caseAHMXB{} population and BBHs simulated with \MESA{}. 
For all but the most massive donor, all \caseAHMXBs{} result in mergers during CE.

\begin{deluxetable*}{ l c c c c c  }
\tablecaption{Fractions \fforward{} of the final outcomes for \caseAHMXBs{}. 
We assume these systems will form a high-spin BH in a HMXBs following a phase of Case-A MT while both stars on the MS. 
From left to right these columns show the fractions of binaries simulated with \COSMIC{} that resulted in BBHs, failed CE, and wide binaries that will not merge within a Hubble time (for simplicity we have combined wide NSBH and wide BBHs systems) 
For models following \citet{belczynski_compact_2008} and \citet{neijssel_effect_2019}, these fractions are illustrated in Figure~\ref{fig:startrack_claeys_fractions} and Figure~\ref{fig:neijssel_claeys_fractions}, respectively.
\label{table:HMXB_outcomes}}
\tablehead{
\colhead{} & & \multicolumn{3}{c}{{\COSMIC{} outcome}} \\
\cline{3-5}
\colhead{Model} & \colhead{$M_{\mathrm{donor}}$} & \colhead{BBH$_{\COSMIC{}}$} & \colhead{Failed CE} & \colhead{Wide binaries} & \colhead{BBH$_{\MESA{}}$}
}
\startdata 
Belczynski et al. (2008) & $25M_{\odot}$ & 0 & 0.49 & 0.52 & 0.12  \\
 & $30M_{\odot}$ & 0 & 0.57 & 0.43  & 0.09 \\
 & $35M_{\odot}$ & 0 & 0.64 & 0.36  & 0.06 \\
 & $40M_{\odot}$ & 0 & 0.77 & 0.23 & 0.05\\
 & $45M_{\odot}$ & 0 & 0.84 & 0.16 & 0.02\\ \hline
Neijssel et al. (2019) & $25M_{\odot}$ & -- & -- & -- & -- \\
 & $30M_{\odot}$ & 0 & 1 & 0 & 0 \\
 & $35M_{\odot}$ & 0 & 1 & 0 & 0 \\
 & $40M_{\odot}$ & 0 & 1 & 0 & 0 \\
 & $45M_{\odot}$ & 0.01 & 0.99 & 0 & 0.01\\ \hline
Claeys et al. (2014) & $25M_{\odot}$ & -- & -- & -- & -- \\
 & $30M_{\odot}$ & 0 & 1 & 0 & 0 \\
 & $35M_{\odot}$ & 0 & 1 & 0 & 0 \\
 & $40M_{\odot}$ & 0 & 1 & 0 & 0 \\
 & $45M_{\odot}$ & 0.01 & 0.99 & 0 & 0 \\
\enddata

\end{deluxetable*}

A summary of the final outcomes for all three models is shown in Table~\ref{table:HMXB_outcomes}. 
The inner four columns correspond to the different final outcomes from the \COSMIC{} simulations. 
The last column corresponds to the fraction of binaries that resulted in BBHs after combining the \COSMIC{} \caseAHMXB{} population with our grids of \MESA{} simulations, BBH$_{\MESA{}}$.

We also assessed whether the values of BBH$_{\MESA{}}$ or BBH$_{\COSMIC{}}$ are affected by different initial binary parameter distributions.
Choosing each initial ZAMS parameter of the binary independently, we found a change of at most $1.8$ in the values of BBH$_{\MESA{}}$ and BBH$_{\COSMIC{}}$ assuming $q_{\mathrm{crit}}$ follows \citet{neijssel_effect_2019}.

\subsection{Fraction of high-spin BBHs}\label{sec:fraction_highspin_BBH_appendix}

Here, we discuss \fbackward{}, the number of BBHs with \caseAHMXB{} progenitors, for the two additional models.

\begin{figure}
\centering
\includegraphics[width=0.46\textwidth]{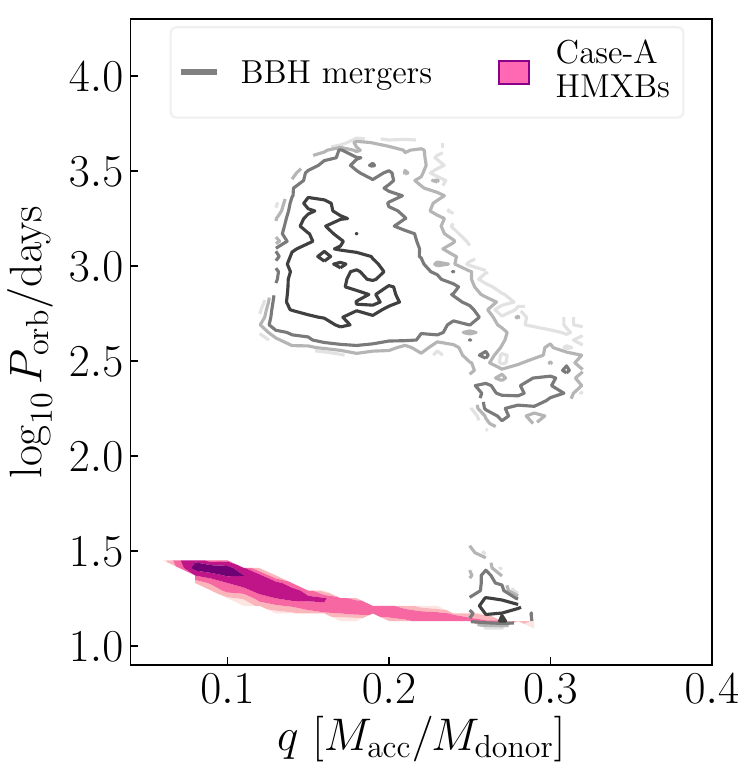}
\caption{Contours showing the population from our \COSMIC{} simulations of all BBHs regardless of their formation path (gray contours) and \caseAHMXBs{} (pink contours) for the model using $q_{\mathrm{crit}}$ following \cite{neijssel_effect_2019} for systems with donor mass $M_{\mathrm{donor}}=(45\pm 2.5)M_{\odot}$.
These populations are shown as a function of mass ratio $q$ and orbital period when the system became a BH--H-rich star. 
The overlapping region corresponds to BBHs that had \caseAHMXBs{} progenitors.} 
\label{fig:neijssel_contour}
\end{figure}

In Figure~\ref{fig:neijssel_contour} we show the \COSMIC{} population of all BBHs regardless of their formation path (gray contours) and all \caseAHMXBs{} (pink contours).
These populations are for BH--H-rich star systems with a donor mass $M_{\mathrm{donor}}=(45\pm 2.5)M_{\odot}$ and $q_{\mathrm{crit}}$ following \citet{neijssel_effect_2019}, as illustrated in the middle panel in Figure~\ref{fig:neijssel_claeys_fractions}. 
Figure~\ref{fig:neijssel_contour} illustrates that these two populations, BBHs and \caseAHMXBs{}, occur in distinct regions in the $\log P_{\mathrm{orb}}$--$q$ parameter space.
The small overlapping region at roughly $q\sim0.26$ and $P_{\mathrm{orb}}\sim 20$ days corresponds to \caseAHMXBs{} that resulted in BBHs.
These systems only comprise a small fraction of parameter space.
Systems with other donor masses have broadly similar results.
Below this donor mass the overlapping region is smaller, and above this donor mass, this region tends to have similar or greater overlap.

\begin{deluxetable}{ c c c c c c c c }
\tablecaption{The fraction \fbackward{} of BBHs with a \caseAHMXB{} progenitor for the three models.
From top to bottom these correspond to 
\citet{belczynski_compact_2008}, \citet{neijssel_effect_2019} and \citet{Claeys2014}, which we list as B+2018, N+2019, and C+2014, respectively.
The top row of each model corresponds to using \COSMIC{} only. 
The second row for each model corresponds to using our grids of BH--H-rich star simulated with \MESA{}.
\label{table:f_backwards}}
\tablehead{
\colhead{} &  & \multicolumn{5}{c}{{Donor}} \\
\cline{3-7}
 \multicolumn{2}{c}{{Model}} & $25M_{\odot}$ & $30M_{\odot}$ & $35M_{\odot}$ & $40M_{\odot}$ & $45M_{\odot}$
}
\startdata 
B+2008 & \COSMIC{}  & 0 & 0 & 0 & 0 & 0 \\
       & \MESA{}    & 0.05 & 0.07 & 0.11 & 0.20 & 0.10\\
N+2019 & \COSMIC{}  & 0 & 0 & 0 & 0 & 0.008 \\
       & \MESA{}    & 0 & 0 & 0 & 0.001 & 0.039 \\
C+2014 & \COSMIC{}  & 0 & 0 & 0 & 0 & 0.005 \\
       & \MESA{}    & 0 & 0 & 0 & 0 & 0 \\
\enddata
\end{deluxetable}

In Table~\ref{table:f_backwards} we show the fraction \fbackward{} of BBHs that had a \caseAHMXB{} progenitor for all our models. 
We show \fbackward{} for systems that we follow the full evolution using only \COSMIC{} and for systems that use our grids of \MESA{} simulations.
Columns in Table~\ref{table:f_backwards} correspond to the different donor mass ranges and rows correspond to the different models. 
These small fractions  indicates that \caseAHMXB{} systems and BBHs likely have little association.

Similar to our results for BBH$_{\MESA{}}$ and BBH$_{\COSMIC{}}$, we also test the robustness of these results when implementing independently distributed initial ZAMS binary parameters compared to a multidimensional joint distribution.   
With an independent distribution, our results for \fbackward{} for the model following $q_{\mathrm{crit}}$ from \cite{neijssel_effect_2019} change by a factor of at most $5$. 
We find a change of a factor of at most $1.8$ for simulations following $q_{\mathrm{crit}}$ from \cite{belczynski_compact_2008}.
Small variations, on the order of $\lesssim5$, in the number of BBHs appear to be in agreement with variations on rates of BBHs due to different initial binary parameters \citep{deMink2015,KlenckiMoe2018}.

\section{Alternative formation scenarios for high-spin BHs in HMXBs}
\label{sec:alternative_formation}

In addition to the Case-A MT scenario adopted here \citep{Qin2019,Valsecchi2010Nature}, several formation channels to form high-spin BHs have been proposed. 
Here we discuss a few alternative channels for forming a high-spin BH as the first born BH in the binary and their possible contributions to the merging BBH population.

One possibility for spinning up BHs in binaries is through accretion. 
A long-lived phase of Eddington-limited accretion can explain the high-spin BHs in low-mass X-ray binaries \citep{Podsiadlowski2003,FragosMcClintock2015}.
In HMXBs, it is thought that the timescale for MT onto the BH is too short for Eddington-limited accretion to substantially spin up the BH \citep{King1999a,FragosMcClintock2015,Mandel2020a}. 
In a case study for the HMXB Cygnus X-1, using simulations ran with \MESA{}, \cite{Qin2022} modeled hypercritical accretion on to a BH, where the mass accretion rate $\dot{M}$ can be a factor of $\sim 10^3$ higher than its Eddington-limited accretion rate $\dot{M}_{\mathrm{Edd}}$. 
They show that a near maximally spinning BH can be formed at these accretion rates under the assumptions of conservative MT and spin-up by accretion from a thin disk.
This resulted in a binary that resembles Cygnus X-1 given its large uncertainties.
Although \cite{Qin2022} did not model the evolution after the formation of this maximally spinning BH, it has been shown that super-Eddington accretion is inefficient at forming merging BBHs \citep{vanSon2020,bavera2021_impact_of_MT,ZevinBavera2022}.
This is because once the BH accretes significant mass and the mass ratio is reversed, conservative MT widens the orbit and prevents a BBH merger within a Hubble time.
As a result, high-spin HMXBs formed via hypercritical accretion will likely not contribute significantly to the population of merging BBHs.
However, in a recent study using \texttt{BPASS}, a population-synthesis code that models the response of the donor star to mass loss \citep{Eldridge2017, StanwayEldridge2018}, \cite{Briel2022} found that super-Eddington accretion can result in binaries with significantly unequal mass ratios when the first BH is formed, enough to enable a BBH merger within a Hubble time. 
Whether these binaries result in a BBH merger or not, it is unclear whether hypercritical or super-Eddington accretion can effectively spin up a BH (\citealt[][Section 1.2]{FragosMcClintock2015}; \citealt[][Section 5.2.3]{vanSon2020}). 
Given these uncertainties we do not consider this scenario in this study. 

In a recent study, \cite{Shao2022} showed that a slow phase of stable Case-A MT lasting $\sim 0.7$ Myr from an $80M_{\odot}$ MS donor onto a $30M_{\odot}$ BH with an initial orbital period of 4 days can form a BBH with a component spin of $\sim0.6$.
This is unlike the Case-A MT studied here, which occurs between two MS stars.
To achieve this, the maximum accretion rate onto the BH was relaxed to $10  \dot{M}_{\mathrm{Edd}}$ \citep{Begelman2002,McKinney2014}.
Although they show that this MT allows for more accretion onto the BH, it is not clear how common the initial conditions required for a slow phase of stable MT are in nature.
Without modeling of the prior evolution that may result in these binaries, and without an informed astrophysical population, it is difficult to determine if these initial condition reflect those of HMXBs or what the contribution of these systems are to the total merging BBH population.   
In \cite{Gallegos-Garcia2021} we simulated MT at $10 \dot{M}_{\mathrm{Edd}}$ for grids of BH--H-rich star binaries with a maximum MS donor mass of $40 M_{\odot}$. 
We found that the BH mass can increase by at least a factor of $1.3$, similar to that shown in \citet{Shao2022}, but only for initial orbital periods $\lesssim 2.5~\mathrm{days}$ when the system is a BH--H-rich star binary.
The contribution of BBHs from this scenario may therefore be similar to the mechanisms mentioned above that invoke accretion rates above the Eddington limit. 
As described for the model implementing hypercritical accretion on to a BH, we do not expect a significant contribution from these channels due to widening of the orbit and also due to possibly strict requirements on initial conditions.

High-spin BHs have also been suggested to form without invoking Roche lobe overflow accretion onto the BH.
New-born BH can be spun-up during a failed or weak SN explosion \citep{Batta2017, Schroder2018}, even if the total angular momentum of the envelope of the SN progenitor is initially zero \citep{Antoni2022}. 
\cite{Batta2017} studied this scenario using three-dimensional smooth particle hydrodynamics simulations for a BH forming in a binary. 
They show how a BH can be spun up by accreting SN fallback material that has been torqued by the companion during a failed SN explosion.
They find that an initially non-spinning BH can reach spins of $\sim$~0.8, but only if the ejected material reaches distances that are comparable to the binary's separation before it is accreted.
Most massive BHs are assumed to form without an explosion \citep{Fryer2012,Ertl2020}, and additionally are expected to have lost their envelope prior to core collapse \citep{Sukhbold2016}, which allows less mass to be accreted by the new-born BH.
Therefore, since our donor stars are massive, we assume this scenario does not play a large role in our populations.

It is still plausible that the spin of more massive BHs can be enhanced during a SN.
\cite{Batta2019} use an analytic formalism to calculate how the resulting mass and spin of a BH from a pre-SN He-star is affected as it accretes shells of stellar material during its direct collapse to a BH.
They show that a rapidly rotating pre-SN He-star can form a BH with high spin values of $> 0.8$ as long as accretion feedback is inefficient.
However, if accretion feedback is strong the expected spin of the BH decreases. 
While this scenario provides a mechanism for forming high-spin BHs in HMXBs, it depends strongly on the rotation rate of the progenitor, which we cannot extract from our simulations. 
As a result, we do not consider this scenario here.

In addition to Case-A MT between two MS stars, \cite{Qin2019} also explored CHE \citep[][]{Mandel2016,Marchant2016,Song2016} as a way to form high-spin BHs in HMXBs.
They found that while this channel can produce high-spin BHs, the orbital periods are too wide compared to observed HMXBs. 
While CHE can still play a role in the formation BBHs with high spin, our goal in this study is to find a scenario that can explain HMXBs with high spin.
We do not consider this scenario and leave it for future work. 

These scenarios for high-spin BHs in HMXBs, including the Case-A MT scenario that forms the \caseAHMXBs{} studied here, all include different assumptions about stellar and binary evolution or SN physics. 
In the context of explaining both high-spin HMXBs and GW observations, we can straightforwardly assess the number of \caseAHMXBs{} in a population and model its subsequent evolution. 
Based on our results from Section~\ref{sec:results}, it appears to satisfy the conditions for HMXBs and merging BBHs.
We leave more detailed analysis of the other scenarios for future work. 

\bibliography{ms}

\begin{thebibliography}{}
\expandafter\ifx\csname natexlab\endcsname\relax\def\natexlab#1{#1}\fi
\providecommand{\url}[1]{\href{#1}{#1}}
\providecommand{\dodoi}[1]{doi:~\href{http://doi.org/#1}{\nolinkurl{#1}}}
\providecommand{\doeprint}[1]{\href{http://ascl.net/#1}{\nolinkurl{http://ascl.net/#1}}}
\providecommand{\doarXiv}[1]{\href{https://arxiv.org/abs/#1}{\nolinkurl{https://arxiv.org/abs/#1}}}

\bibitem[{{Abbott} {et~al.}(2021{\natexlab{a}}){Abbott}, {Abbott}, {Acernese},
  {Ackley}, {Adams}, {Adhikari}, {Adhikari}, \& et~al.}]{2021LIGO_GWTC-3}
{Abbott}, R., {Abbott}, T.~D., {Acernese}, F., {et~al.} 2021{\natexlab{a}},
  arXiv e-prints, arXiv:2111.03606.
\newblock \doarXiv{2111.03606}

\bibitem[{{Abbott} {et~al.}(2021{\natexlab{b}}){Abbott}, {Abbott}, {Acernese},
  {Ackley}, {Adams}, {Adhikari}, {Adhikari}, {Adya}, \&
  et~al.}]{2021LIGO_GWTC-2.1}
---. 2021{\natexlab{b}}, arXiv e-prints, arXiv:2108.01045.
\newblock \doarXiv{2108.01045}

\bibitem[{{Abbott} {et~al.}(2021{\natexlab{c}}){Abbott}, {Abbott}, {Acernese},
  {Ackley}, {Adams}, {Adhikari}, {Adhikari}, {Adya}, {Affeldt}, {Agarwal},
  {Agathos}, {Agatsuma}, {Aggarwal}, {Aguiar}, {Aiello}, {Ain}, {Ajith},
  {Akutsu}, {Albanesi}, {Allocca}, {Altin}, {Amato}, {Anand}, {Anand},
  {Ananyeva}, {Anderson}, {Anderson}, {Ando}, {Andrade}, {Andres},
  {Andri{\'c}}, {Angelova}, {Ansoldi}, {Antelis}, {Antier}, {Antonini},
  {Appert}, {Arai}, {Arai}, {Arai}, {Araki}, {Araya}, {Araya}, {Areeda},
  {Ar{\`e}ne}, {Aritomi}, {Arnaud}, {Aronson}, {Arun}, {Asada}, {Asali},
  {Ashton}, {Aso}, {Assiduo}, {Aston}, {Astone}, {Aubin}, {Austin}, {Babak},
  {Badaracco}, {Bader}, {Badger}, {Bae}, {Bae}, {Baer}, {Bagnasco}, {Bai},
  {Baiotti}, {Baird}, {Bajpai}, {Ball}, {Ballardin}, {Ballmer}, {Balsamo},
  {Baltus}, {Banagiri}, {Bankar}, {Barayoga}, {Barbieri}, {Barish}, {Barker},
  {Barneo}, {Barone}, {Barr}, {Barsotti}, {Barsuglia}, {Barta}, {Bartlett},
  {Barton}, {Bartos}, {Bassiri}, {Basti}, {Bawaj}, {Bayley}, {Baylor},
  {Bazzan}, {B{\'e}csy}, {Bedakihale}, {Bejger}, {Belahcene}, {Benedetto},
  {Beniwal}, {Bennett}, {Bentley}, {BenYaala}, {Bergamin}, {Berger},
  {Bernuzzi}, {Berry}, {Bersanetti}, {Bertolini}, {Betzwieser}, {Beveridge},
  {Bhandare}, {Bhardwaj}, {Bhattacharjee}, {Bhaumik}, {Bilenko}, {Billingsley},
  {Bini}, {Birney}, {Birnholtz}, {Biscans}, {Bischi}, {Biscoveanu}, {Bisht},
  {Biswas}, {Bitossi}, {Bizouard}, {Blackburn}, {Blair}, {Blair}, {Blair},
  {Bobba}, {Bode}, {Boer}, {Bogaert}, {Boldrini}, {Bonavena}, {Bondu},
  {Bonilla}, {Bonnand}, {Booker}, {Boom}, {Bork}, {Boschi}, {Bose}, {Bose},
  {Bossilkov}, {Boudart}, {Bouffanais}, {Bozzi}, {Bradaschia}, {Brady},
  {Bramley}, {Branch}, {Branchesi}, {Brau}, {Breschi}, {Briant}, {Briggs},
  {Brillet}, {Brinkmann}, {Brockill}, {Brooks}, {Brooks}, {Brown}, {Brunett},
  {Bruno}, {Bruntz}, {Bryant}, {Bulik}, {Bulten}, {Buonanno}, {Buscicchio},
  {Buskulic}, {Buy}, {Byer}, {Cadonati}, {Cagnoli}, {Cahillane}, {Calder{\'o}n
  Bustillo}, {Callaghan}, {Callister}, {Calloni}, {Cameron}, {Camp}, {Canepa},
  {Canevarolo}, {Cannavacciuolo}, {Cannon}, {Cao}, {Cao}, {Capocasa}, {Capote},
  {Carapella}, {Carbognani}, {Carlin}, {Carney}, {Carpinelli}, {Carrillo},
  {Carullo}, {Carver}, {Casanueva Diaz}, {Casentini}, {Castaldi}, {Caudill},
  {Cavagli{\`a}}, {Cavalier}, {Cavalieri}, {Ceasar}, {Cella},
  {Cerd{\'a}-Dur{\'a}n}, {Cesarini}, {Chaibi}, {Chakravarti}, {Chalathadka
  Subrahmanya}, {Champion}, {Chan}, {Chan}, {Chan}, {Chan}, {Chan}, {Chandra},
  {Chanial}, {Chao}, {Charlton}, {Chase}, {Chassande-Mottin}, {Chatterjee},
  {Chatterjee}, {Chatterjee}, {Chaturvedi}, {Chaty}, {Chatziioannou}, {Chen},
  {Chen}, {Chen}, {Chen}, {Chen}, {Chen}, {Chen}, {Chen}, {Cheng}, {Cheong},
  {Cheung}, {Chia}, {Chiadini}, {Chiang}, {Chiarini}, {Chierici}, {Chincarini},
  {Chiofalo}, {Chiummo}, {Cho}, {Cho}, {Choudhary}, {Choudhary}, {Christensen},
  {Chu}, {Chu}, {Chu}, {Chua}, {Chung}, {Ciani}, {Ciecielag}, {Cie{\'s}lar},
  {Cifaldi}, {Ciobanu}, {Ciolfi}, {Cipriano}, {Cirone}, {Clara}, {Clark},
  {Clark}, {Clarke}, {Clearwater}, {Clesse}, {Cleva}, {Coccia}, {Codazzo},
  {Cohadon}, {Cohen}, {Cohen}, {Colleoni}, {Collette}, {Colombo}, {Colpi},
  {Compton}, {Constancio}, {Conti}, {Cooper}, {Corban}, {Corbitt},
  {Cordero-Carri{\'o}n}, {Corezzi}, {Corley}, {Cornish}, {Corre}, {Corsi},
  {Cortese}, {Costa}, {Cotesta}, {Coughlin}, {Coulon}, {Countryman}, {Cousins},
  {Couvares}, {Coward}, {Cowart}, {Coyne}, {Coyne}, {Creighton}, {Creighton},
  {Criswell}, {Croquette}, {Crowder}, {Cudell}, {Cullen}, {Cumming},
  {Cummings}, {Cunningham}, {Cuoco}, {Cury{\l}o}, {Dabadie}, {Dal Canton},
  {Dall'Osso}, {D{\'a}lya}, {Dana}, {DaneshgaranBajastani}, {D'Angelo},
  {Danilishin}, {D'Antonio}, {Danzmann}, {Darsow-Fromm}, {Dasgupta}, {Datrier},
  {Datta}, {Dattilo}, {Dave}, {Davier}, {Davies}, {Davis}, {Davis}, {Daw},
  {Dean}, {DeBra}, {Deenadayalan}, {Degallaix}, {De Laurentis},
  {Del{\'e}glise}, {Del Favero}, {De Lillo}, {De Lillo}, {Del Pozzo},
  {DeMarchi}, {De Matteis}, {D'Emilio}, {Demos}, {Dent}, {Depasse}, {De
  Pietri}, {De Rosa}, {De Rossi}, {DeSalvo}, {De Simone}, {Dhurandhar},
  {D{\'\i}az}, {Diaz-Ortiz}, {Didio}, {Dietrich}, {Di Fiore}, {Di Fronzo}, {Di
  Giorgio}, {Di Giovanni}, {Di Giovanni}, {Di Girolamo}, {Di Lieto}, {Ding},
  {Di Pace}, {Di Palma}, {Di Renzo}, {Divakarla}, {Dmitriev}, {Doctor},
  {D'Onofrio}, {Donovan}, {Dooley}, {Doravari}, {Dorrington}, {Drago},
  {Driggers}, {Drori}, {Ducoin}, {Dupej}, {Durante}, {D'Urso}, {Duverne},
  {Dwyer}, {Eassa}, {Easter}, {Ebersold}, {Eckhardt}, {Eddolls}, {Edelman},
  {Edo}, {Edy}, {Effler}, {Eguchi}, {Eichholz}, {Eikenberry}, {Eisenmann},
  {Eisenstein}, {Ejlli}, {Engelby}, {Enomoto}, {Errico}, {Essick},
  {Estell{\'e}s}, {Estevez}, {Etienne}, {Etzel}, {Evans}, {Evans}, {Ewing},
  {Fafone}, {Fair}, {Fairhurst}, {Farah}, {Farinon}, {Farr}, {Farr}, {Farrow},
  {Fauchon-Jones}, {Favaro}, {Favata}, {Fays}, {Fazio}, {Feicht}, {Fejer},
  {Fenyvesi}, {Ferguson}, {Fernandez-Galiana}, {Ferrante}, {Ferreira},
  {Fidecaro}, {Figura}, {Fiori}, {Fishbach}, {Fisher}, {Fittipaldi}, {Fiumara},
  {Flaminio}, {Floden}, {Fong}, {Font}, {Fornal}, {Forsyth}, {Franke},
  {Frasca}, {Frasconi}, {Frederick}, {Freed}, {Frei}, {Freise}, {Frey},
  {Fritschel}, {Frolov}, {Fronz{\'e}}, {Fujii}, {Fujikawa}, {Fukunaga},
  {Fukushima}, {Fulda}, {Fyffe}, {Gabbard}, {Gadre}, {Gair}, {Gais},
  {Galaudage}, {Gamba}, {Ganapathy}, {Ganguly}, {Gao}, {Gaonkar}, {Garaventa},
  {Garc{\'\i}a-N{\'u}{\~n}ez}, {Garc{\'\i}a-Quir{\'o}s}, {Garufi}, {Gateley},
  {Gaudio}, {Gayathri}, {Ge}, {Gemme}, {Gennai}, {George}, {Gerberding},
  {Gergely}, {Gewecke}, {Ghonge}, {Ghosh}, {Ghosh}, {Ghosh}, {Ghosh},
  {Giacomazzo}, {Giacoppo}, {Giaime}, {Giardina}, {Gibson}, {Gier}, {Giesler},
  {Giri}, {Gissi}, {Glanzer}, {Gleckl}, {Godwin}, {Goetz}, {Goetz}, {Gohlke},
  {Golomb}, {Goncharov}, {Gonz{\'a}lez}, {Gopakumar}, {Gosselin}, {Gouaty},
  {Gould}, {Grace}, {Grado}, {Granata}, {Granata}, {Grant}, {Gras}, {Grassia},
  {Gray}, {Gray}, {Greco}, {Green}, {Green}, {Gretarsson}, {Gretarsson},
  {Griffith}, {Griffiths}, {Griggs}, {Grignani}, {Grimaldi}, {Grimm}, {Grote},
  {Grunewald}, {Gruning}, {Guerra}, {Guidi}, {Guimaraes}, {Guix{\'e}},
  {Gulati}, {Guo}, {Guo}, {Gupta}, {Gupta}, {Gupta}, {Gustafson}, {Gustafson},
  {Guzman}, {Ha}, {Haegel}, {Hagiwara}, {Haino}, {Halim}, {Hall}, {Hamilton},
  {Hammond}, {Han}, {Haney}, {Hanks}, {Hanna}, {Hannam}, {Hannuksela},
  {Hansen}, {Hansen}, {Hanson}, {Harder}, {Hardwick}, {Haris}, {Harms},
  {Harry}, {Harry}, {Hartwig}, {Hasegawa}, {Haskell}, {Hasskew}, {Haster},
  {Hattori}, {Haughian}, {Hayakawa}, {Hayama}, {Hayes}, {Healy}, {Heidmann},
  {Heidt}, {Heintze}, {Heinze}, {Heinzel}, {Heitmann}, {Hellman}, {Hello},
  {Helmling-Cornell}, {Hemming}, {Hendry}, {Heng}, {Hennes}, {Hennig},
  {Hennig}, {Hernandez}, {Hernandez Vivanco}, {Heurs}, {Hild}, {Hill},
  {Himemoto}, {Hines}, {Hiranuma}, {Hirata}, {Hirose}, {Hochheim}, {Hofman},
  {Hohmann}, {Holcomb}, {Holland}, {Hollows}, {Holmes}, {Holt}, {Holz}, {Hong},
  {Hopkins}, {Hough}, {Hourihane}, {Howell}, {Hoy}, {Hoyland}, {Hreibi},
  {Hsieh}, {Hsu}, {Huang}, {Huang}, {Huang}, {Huang}, {Huang}, {Huang},
  {H{\"u}bner}, {Huddart}, {Hughey}, {Hui}, {Hui}, {Husa}, {Huttner},
  {Huxford}, {Huynh-Dinh}, {Ide}, {Idzkowski}, {Iess}, {Ikenoue}, {Imam},
  {Inayoshi}, {Ingram}, {Inoue}, {Ioka}, {Isi}, {Isleif}, {Ito}, {Itoh},
  {Iyer}, {Izumi}, {JaberianHamedan}, {Jacqmin}, {Jadhav}, {Jadhav}, {James},
  {Jan}, {Jani}, {Janquart}, {Janssens}, {Janthalur}, {Jaranowski}, {Jariwala},
  {Jaume}, {Jenkins}, {Jenner}, {Jeon}, {Jeunon}, {Jia}, {Jin}, {Johns},
  {Jones}, {Jones}, {Jones}, {Jones}, {Jones}, {Jonker}, {Ju}, {Jung}, {Jung},
  {Junker}, {Juste}, {Kaihotsu}, {Kajita}, {Kakizaki}, {Kalaghatgi},
  {Kalogera}, {Kamai}, {Kamiizumi}, {Kanda}, {Kandhasamy}, {Kang}, {Kanner},
  {Kao}, {Kapadia}, {Kapasi}, {Karat}, {Karathanasis}, {Karki}, {Kashyap},
  {Kasprzack}, {Kastaun}, {Katsanevas}, {Katsavounidis}, {Katzman}, {Kaur},
  {Kawabe}, {Kawaguchi}, {Kawai}, {Kawasaki}, {K{\'e}f{\'e}lian}, {Keitel},
  {Key}, {Khadka}, {Khalili}, {Khan}, {Khazanov}, {Khetan}, {Khursheed},
  {Kijbunchoo}, {Kim}, {Kim}, {Kim}, {Kim}, {Kim}, {Kim}, {Kimball}, {Kimura},
  {Kinley-Hanlon}, {Kirchhoff}, {Kissel}, {Kita}, {Kitazawa}, {Kleybolte},
  {Klimenko}, {Knee}, {Knowles}, {Knyazev}, {Koch}, {Koekoek}, {Kojima},
  {Kokeyama}, {Koley}, {Kolitsidou}, {Kolstein}, {Komori}, {Kondrashov},
  {Kong}, {Kontos}, {Koper}, {Korobko}, {Kotake}, {Kovalam}, {Kozak},
  {Kozakai}, {Kozu}, {Kringel}, {Krishnendu}, {Kr{\'o}lak}, {Kuehn}, {Kuei},
  {Kuijer}, {Kumar}, {Kumar}, {Kumar}, {Kumar}, {Kume}, {Kuns}, {Kuo}, {Kuo},
  {Kuromiya}, {Kuroyanagi}, {Kusayanagi}, {Kuwahara}, {Kwak}, {Lagabbe},
  {Laghi}, {Lalande}, {Lam}, {Lamberts}, {Landry}, {Landry}, {Lane}, {Lang},
  {Lange}, {Lantz}, {La Rosa}, {Lartaux-Vollard}, {Lasky}, {Laxen},
  {Lazzarini}, {Lazzaro}, {Leaci}, {Leavey}, {Lecoeuche}, {Lee}, {Lee}, {Lee},
  {Lee}, {Lee}, {Lee}, {Lehmann}, {Lema{\^\i}tre}, {Leonardi}, {Leroy},
  {Letendre}, {Levesque}, {Levin}, {Leviton}, {Leyde}, {Li}, {Li}, {Li}, {Li},
  {Li}, {Li}, {Lin}, {Lin}, {Lin}, {Lin}, {Lin}, {Linde}, {Linker}, {Linley},
  {Littenberg}, {Liu}, {Liu}, {Liu}, {Liu}, {Llamas}, {Llorens-Monteagudo},
  {Lo}, {Lockwood}, {London}, {Longo}, {Lopez}, {Lopez Portilla}, {Lorenzini},
  {Loriette}, {Lormand}, {Losurdo}, {Lott}, {Lough}, {Lousto}, {Lovelace},
  {Lucaccioni}, {L{\"u}ck}, {Lumaca}, {Lundgren}, {Luo}, {Lynam}, {Macas},
  {MacInnis}, {Macleod}, {MacMillan}, {Macquet}, {Maga{\~n}a Hernandez},
  {Magazz{\`u}}, {Magee}, {Maggiore}, {Magnozzi}, {Mahesh}, {Majorana},
  {Makarem}, {Maksimovic}, {Maliakal}, {Malik}, {Man}, {Mandic}, {Mangano},
  {Mango}, {Mansell}, {Manske}, {Mantovani}, {Mapelli}, {Marchesoni},
  {Marchio}, {Marion}, {Mark}, {M{\'a}rka}, {M{\'a}rka}, {Markakis},
  {Markosyan}, {Markowitz}, {Maros}, {Marquina}, {Marsat}, {Martelli},
  {Martin}, {Martin}, {Martinez}, {Martinez}, {Martinez}, {Martinovic},
  {Martynov}, {Marx}, {Masalehdan}, {Mason}, {Massera}, {Masserot},
  {Massinger}, {Masso-Reid}, {Mastrogiovanni}, {Matas}, {Mateu-Lucena},
  {Matichard}, {Matiushechkina}, {Mavalvala}, {McCann}, {McCarthy},
  {McClelland}, {McClincy}, {McCormick}, {McCuller}, {McGhee}, {McGuire},
  {McIsaac}, {McIver}, {McRae}, {McWilliams}, {Meacher}, {Mehmet}, {Mehta},
  {Meijer}, {Melatos}, {Melchor}, {Mendell}, {Menendez-Vazquez}, {Menoni},
  {Mercer}, {Mereni}, {Merfeld}, {Merilh}, {Merritt}, {Merzougui}, {Meshkov},
  {Messenger}, {Messick}, {Meyers}, {Meylahn}, {Mhaske}, {Miani}, {Miao},
  {Michaloliakos}, {Michel}, {Michimura}, {Middleton}, {Milano}, {Miller},
  {Miller}, {Miller}, {Miller}, {Millhouse}, {Mills}, {Milotti}, {Minazzoli},
  {Minenkov}, {Mio}, {Mir}, {Miravet-Ten{\'e}s}, {Mishra}, {Mishra}, {Mistry},
  {Mitra}, {Mitrofanov}, {Mitselmakher}, {Mittleman}, {Miyakawa}, {Miyamoto},
  {Miyazaki}, {Miyo}, {Miyoki}, {Mo}, {Moguel}, {Mogushi}, {Mohapatra},
  {Mohite}, {Molina}, {Molina-Ruiz}, {Mondin}, {Montani}, {Moore}, {Moraru},
  {Morawski}, {More}, {Moreno}, {Moreno}, {Mori}, {Morisaki}, {Moriwaki},
  {Mours}, {Mow-Lowry}, {Mozzon}, {Muciaccia}, {Mukherjee}, {Mukherjee},
  {Mukherjee}, {Mukherjee}, {Mukherjee}, {Mukund}, {Mullavey}, {Munch},
  {Mu{\~n}iz}, {Murray}, {Musenich}, {Muusse}, {Nadji}, {Nagano}, {Nagano},
  {Nagar}, {Nakamura}, {Nakano}, {Nakano}, {Nakashima}, {Nakayama}, {Napolano},
  {Nardecchia}, {Narikawa}, {Naticchioni}, {Nayak}, {Nayak}, {Negishi}, {Neil},
  {Neilson}, {Nelemans}, {Nelson}, {Nery}, {Neubauer}, {Neunzert}, {Ng}, {Ng},
  {Nguyen}, {Nguyen}, {Nguyen}, {Nguyen Quynh}, {Ni}, {Nichols}, {Nishizawa},
  {Nissanke}, {Nitoglia}, {Nocera}, {Norman}, {North}, {Nozaki}, {Nuttall},
  {Oberling}, {O'Brien}, {Obuchi}, {O'Dell}, {Oelker}, {Ogaki}, {Oganesyan},
  {Oh}, {Oh}, {Oh}, {Ohashi}, {Ohishi}, {Ohkawa}, {Ohme}, {Ohta}, {Okada},
  {Okutani}, {Okutomi}, {Olivetto}, {Oohara}, {Ooi}, {Oram}, {O'Reilly},
  {Ormiston}, {Ormsby}, {Ortega}, {O'Shaughnessy}, {O'Shea}, {Oshino},
  {Ossokine}, {Osthelder}, {Otabe}, {Ottaway}, {Overmier}, {Pace}, {Pagano},
  {Page}, {Pagliaroli}, {Pai}, {Pai}, {Palamos}, {Palashov}, {Palomba}, {Pan},
  {Pan}, {Panda}, {Pang}, {Pang}, {Pankow}, {Pannarale}, {Pant}, {Panther},
  {Paoletti}, {Paoli}, {Paolone}, {Parisi}, {Park}, {Park}, {Parker},
  {Pascucci}, {Pasqualetti}, {Passaquieti}, {Passuello}, {Patel}, {Pathak},
  {Patricelli}, {Patron}, {Paul}, {Payne}, {Pedraza}, {Pegoraro}, {Pele},
  {Pe{\~n}a Arellano}, {Penn}, {Perego}, {Pereira}, {Pereira}, {Perez},
  {P{\'e}rigois}, {Perkins}, {Perreca}, {Perri{\`e}s}, {Petermann},
  {Petterson}, {Pfeiffer}, {Pham}, {Phukon}, {Piccinni}, {Pichot},
  {Piendibene}, {Piergiovanni}, {Pierini}, {Pierro}, {Pillant}, {Pillas},
  {Pilo}, {Pinard}, {Pinto}, {Pinto}, {Piotrzkowski}, {Pirello}, {Pitkin},
  {Placidi}, {Planas}, {Plastino}, {Pluchar}, {Poggiani}, {Polini}, {Pong},
  {Ponrathnam}, {Popolizio}, {Porter}, {Poulton}, {Powell}, {Pracchia},
  {Pradier}, {Prajapati}, {Prasai}, {Prasanna}, {Pratten}, {Principe}, {Prodi},
  {Prokhorov}, {Prosposito}, {Prudenzi}, {Puecher}, {Punturo}, {Puosi},
  {Puppo}, {P{\"u}rrer}, {Qi}, {Quetschke}, {Quitzow-James}, {Raab},
  {Raaijmakers}, {Radkins}, {Radulesco}, {Raffai}, {Rail}, {Raja}, {Rajan},
  {Ramirez}, {Ramirez}, {Ramos-Buades}, {Rana}, {Rapagnani}, {Rapol}, {Ray},
  {Raymond}, {Raza}, {Razzano}, {Read}, {Rees}, {Regimbau}, {Rei}, {Reid},
  {Reid}, {Reitze}, {Relton}, {Renzini}, {Rettegno}, {Rezac}, {Ricci},
  {Richards}, {Richardson}, {Richardson}, {Riemenschneider}, {Riles},
  {Rinaldi}, {Rink}, {Rizzo}, {Robertson}, {Robie}, {Robinet}, {Rocchi},
  {Rodriguez}, {Rolland}, {Rollins}, {Romanelli}, {Romano}, {Romel},
  {Romero-Rodr{\'\i}guez}, {Romero-Shaw}, {Romie}, {Ronchini}, {Rosa}, {Rose},
  {Rosi{\'n}ska}, {Ross}, {Rowan}, {Rowlinson}, {Roy}, {Roy}, {Roy}, {Rozza},
  {Ruggi}, {Ryan}, {Sachdev}, {Sadecki}, {Sadiq}, {Sago}, {Saito}, {Saito},
  {Sakai}, {Sakai}, {Sakellariadou}, {Sakuno}, {Salafia}, {Salconi}, {Saleem},
  {Salemi}, {Samajdar}, {Sanchez}, {Sanchez}, {Sanchez}, {Sanchis-Gual},
  {Sanders}, {Sanuy}, {Saravanan}, {Sarin}, {Sassolas}, {Satari},
  {Sathyaprakash}, {Sato}, {Sato}, {Sauter}, {Savage}, {Sawada}, {Sawant},
  {Sawant}, {Sayah}, {Schaetzl}, {Scheel}, {Scheuer}, {Schiworski}, {Schmidt},
  {Schmidt}, {Schnabel}, {Schneewind}, {Schofield}, {Sch{\"o}nbeck}, {Schulte},
  {Schutz}, {Schwartz}, {Scott}, {Scott}, {Seglar-Arroyo}, {Sekiguchi},
  {Sekiguchi}, {Sellers}, {Sengupta}, {Sentenac}, {Seo}, {Sequino}, {Sergeev},
  {Setyawati}, {Shaffer}, {Shahriar}, {Shams}, {Shao}, {Sharma}, {Sharma},
  {Shawhan}, {Shcheblanov}, {Shibagaki}, {Shikauchi}, {Shimizu}, {Shimoda},
  {Shimode}, {Shinkai}, {Shishido}, {Shoda}, {Shoemaker}, {Shoemaker},
  {ShyamSundar}, {Sieniawska}, {Sigg}, {Singer}, {Singh}, {Singh}, {Singha},
  {Sintes}, {Sipala}, {Skliris}, {Slagmolen}, {Slaven-Blair}, {Smetana},
  {Smith}, {Smith}, {Soldateschi}, {Somala}, {Somiya}, {Son}, {Soni}, {Soni},
  {Sordini}, {Sorrentino}, {Sorrentino}, {Sotani}, {Soulard}, {Souradeep},
  {Sowell}, {Spagnuolo}, {Spencer}, {Spera}, {Srinivasan}, {Srivastava},
  {Srivastava}, {Staats}, {Stachie}, {Steer}, {Steinlechner}, {Steinlechner},
  {Stops}, {Stover}, {Strain}, {Strang}, {Stratta}, {Strunk}, {Sturani},
  {Stuver}, {Sudhagar}, {Sudhir}, {Sugimoto}, {Suh}, {Summerscales}, {Sun},
  {Sun}, {Sunil}, {Sur}, {Suresh}, {Sutton}, {Suzuki}, {Suzuki}, {Swinkels},
  {Szczepa{\'n}czyk}, {Szewczyk}, {Tacca}, {Tagoshi}, {Tait}, {Takahashi},
  {Takahashi}, {Takamori}, {Takano}, {Takeda}, {Takeda}, {Talbot}, {Talbot},
  {Tanaka}, {Tanaka}, {Tanaka}, {Tanaka}, {Tanaka}, {Tanasijczuk}, {Tanioka},
  {Tanner}, {Tao}, {Tao}, {Tapia San Mart{\'\i}n}, {Taranto}, {Tasson},
  {Telada}, {Tenorio}, {Terhune}, {Terkowski}, {Thirugnanasambandam}, {Thomas},
  {Thomas}, {Thompson}, {Thondapu}, {Thorne}, {Thrane}, {Tiwari}, {Tiwari},
  {Tiwari}, {Toivonen}, {Toland}, {Tolley}, {Tomaru}, {Tomigami}, {Tomura},
  {Tonelli}, {Torres-Forn{\'e}}, {Torrie}, {Tosta e Melo}, {T{\"o}yr{\"a}},
  {Trapananti}, {Travasso}, {Traylor}, {Trevor}, {Tringali}, {Tripathee},
  {Troiano}, {Trovato}, {Trozzo}, {Trudeau}, {Tsai}, {Tsai}, {Tsang}, {Tsang},
  {Tsao}, {Tse}, {Tso}, {Tsubono}, {Tsuchida}, {Tsukada}, {Tsuna}, {Tsutsui},
  {Tsuzuki}, {Turbang}, {Turconi}, {Tuyenbayev}, {Ubhi}, {Uchikata},
  {Uchiyama}, {Udall}, {Ueda}, {Uehara}, {Ueno}, {Ueshima}, {Unnikrishnan},
  {Uraguchi}, {Urban}, {Ushiba}, {Utina}, {Vahlbruch}, {Vajente}, {Vajpeyi},
  {Valdes}, {Valentini}, {Valsan}, {van Bakel}, {van Beuzekom}, {van den
  Brand}, {Van Den Broeck}, {Vander-Hyde}, {van der Schaaf}, {van Heijningen},
  {Vanosky}, {van Putten}, {van Remortel}, {Vardaro}, {Vargas}, {Varma},
  {Vas{\'u}th}, {Vecchio}, {Vedovato}, {Veitch}, {Veitch}, {Venneberg},
  {Venugopalan}, {Verkindt}, {Verma}, {Verma}, {Veske}, {Vetrano},
  {Vicer{\'e}}, {Vidyant}, {Viets}, {Vijaykumar}, {Villa-Ortega}, {Vinet},
  {Virtuoso}, {Vitale}, {Vo}, {Vocca}, {von Reis}, {von Wrangel}, {Vorvick},
  {Vyatchanin}, {Wade}, {Wade}, {Wagner}, {Walet}, {Walker}, {Wallace},
  {Wallace}, {Walsh}, {Wang}, {Wang}, {Wang}, {Ward}, {Warner}, {Was},
  {Washimi}, {Washington}, {Watchi}, {Weaver}, {Webster}, {Weinert},
  {Weinstein}, {Weiss}, {Weller}, {Wellmann}, {Wen}, {We{\ss}els}, {Wette},
  {Whelan}, {White}, {Whiting}, {Whittle}, {Wilken}, {Williams}, {Williams},
  {Williamson}, {Willis}, {Willke}, {Wilson}, {Winkler}, {Wipf}, {Wlodarczyk},
  {Woan}, {Woehler}, {Wofford}, {Wong}, {Wu}, {Wu}, {Wu}, {Wu}, {Wysocki},
  {Xiao}, {Xu}, {Yamada}, {Yamamoto}, {Yamamoto}, {Yamamoto}, {Yamamoto},
  {Yamashita}, {Yamazaki}, {Yang}, {Yang}, {Yang}, {Yang}, {Yang}, {Yap},
  {Yeeles}, {Yelikar}, {Ying}, {Yokogawa}, {Yokoyama}, {Yokozawa}, {Yoo},
  {Yoshioka}, {Yu}, {Yu}, {Yuzurihara}, {Zadro{\.z}ny}, {Zanolin}, {Zeidler},
  {Zelenova}, {Zendri}, {Zevin}, {Zhan}, {Zhang}, {Zhang}, {Zhang}, {Zhang},
  {Zhang}, {Zhao}, {Zhao}, {Zhao}, {Zhao}, {Zhou}, {Zhou}, {Zhu}, {Zhu},
  {Zimmerman}, {Zlochower}, {Zucker}, \& {Zweizig}}]{2021LIGO_population}
---. 2021{\natexlab{c}}, arXiv e-prints, arXiv:2111.03634.
\newblock \doarXiv{2111.03634}

\bibitem[{{Abbott} {et~al.}(2021{\natexlab{d}}){Abbott}, {Abbott}, {Abraham},
  {Acernese}, {Ackley}, {Adams}, {Adams}, {Adhikari}, {Adya}, {Affeldt}, \&
  et~al.}]{LIGO2021_population_properties}
{Abbott}, R., {Abbott}, T.~D., {Abraham}, S., {et~al.} 2021{\natexlab{d}},
  \apjl, 913, L7, \dodoi{10.3847/2041-8213/abe949}

\bibitem[{{Abbott} {et~al.}(2021{\natexlab{e}}){Abbott}, {Abbott}, {Abraham},
  {Acernese}, {Ackley}, {Adams}, {Adams}, {Adhikari}, {Adya}, {Affeldt}, \&
  et~al.}]{Abbott2020}
---. 2021{\natexlab{e}}, PhRvX, 11, 021053, \dodoi{10.1103/PhysRevX.11.021053}

\bibitem[{{Ade} {et~al.}(2016){Ade}, {Aghanim}, {Arnaud}, {Ashdown}, {Aumont},
  {Baccigalupi}, {Banday}, {Barreiro}, {Bartlett}, {Bartolo}, {Battaner},
  {Battye}, {Benabed}, {Beno{\^\i}t}, {Benoit-L{\'e}vy}, {Bernard},
  {Bersanelli}, {Bielewicz}, {Bock}, {Bonaldi}, {Bonavera}, {Bond}, {Borrill},
  {Bouchet}, {Boulanger}, {Bucher}, {Burigana}, {Butler}, {Calabrese},
  {Cardoso}, {Catalano}, {Challinor}, {Chamballu}, {Chary}, {Chiang}, {Chluba},
  {Christensen}, {Church}, {Clements}, {Colombi}, {Colombo}, {Combet},
  {Coulais}, {Crill}, {Curto}, {Cuttaia}, {Danese}, {Davies}, {Davis}, {de
  Bernardis}, {de Rosa}, {de Zotti}, {Delabrouille}, {D{\'e}sert}, {Di
  Valentino}, {Dickinson}, {Diego}, {Dolag}, {Dole}, {Donzelli}, {Dor{\'e}},
  {Douspis}, {Ducout}, {Dunkley}, {Dupac}, {Efstathiou}, {Elsner},
  {En{\ss}lin}, {Eriksen}, {Farhang}, {Fergusson}, {Finelli}, {Forni},
  {Frailis}, {Fraisse}, {Franceschi}, {Frejsel}, {Galeotta}, {Galli}, {Ganga},
  {Gauthier}, {Gerbino}, {Ghosh}, {Giard}, {Giraud-H{\'e}raud}, {Giusarma},
  {Gjerl{\o}w}, {Gonz{\'a}lez-Nuevo}, {G{\'o}rski}, {Gratton}, {Gregorio},
  {Gruppuso}, {Gudmundsson}, {Hamann}, {Hansen}, {Hanson}, {Harrison}, {Helou},
  {Henrot-Versill{\'e}}, {Hern{\'a}ndez-Monteagudo}, {Herranz}, {Hildebrandt},
  {Hivon}, {Hobson}, {Holmes}, {Hornstrup}, {Hovest}, {Huang}, {Huffenberger},
  {Hurier}, {Jaffe}, {Jaffe}, {Jones}, {Juvela}, {Keih{\"a}nen}, {Keskitalo},
  {Kisner}, {Kneissl}, {Knoche}, {Knox}, {Kunz}, {Kurki-Suonio}, {Lagache},
  {L{\"a}hteenm{\"a}ki}, {Lamarre}, {Lasenby}, {Lattanzi}, {Lawrence}, {Leahy},
  {Leonardi}, {Lesgourgues}, {Levrier}, {Lewis}, {Liguori}, {Lilje},
  {Linden-V{\o}rnle}, {L{\'o}pez-Caniego}, {Lubin}, {Mac{\'\i}as-P{\'e}rez},
  {Maggio}, {Maino}, {Mandolesi}, {Mangilli}, {Marchini}, {Maris}, {Martin},
  {Martinelli}, {Mart{\'\i}nez-Gonz{\'a}lez}, {Masi}, {Matarrese}, {McGehee},
  {Meinhold}, {Melchiorri}, {Melin}, {Mendes}, {Mennella}, {Migliaccio},
  {Millea}, {Mitra}, {Miville-Desch{\^e}nes}, {Moneti}, {Montier}, {Morgante},
  {Mortlock}, {Moss}, {Munshi}, {Murphy}, {Naselsky}, {Nati}, {Natoli},
  {Netterfield}, {N{\o}rgaard-Nielsen}, {Noviello}, {Novikov}, {Novikov},
  {Oxborrow}, {Paci}, {Pagano}, {Pajot}, {Paladini}, {Paoletti}, {Partridge},
  {Pasian}, {Patanchon}, {Pearson}, {Perdereau}, {Perotto}, {Perrotta},
  {Pettorino}, {Piacentini}, {Piat}, {Pierpaoli}, {Pietrobon}, {Plaszczynski},
  {Pointecouteau}, {Polenta}, {Popa}, {Pratt}, {Pr{\'e}zeau}, {Prunet},
  {Puget}, {Rachen}, {Reach}, {Rebolo}, {Reinecke}, {Remazeilles}, {Renault},
  {Renzi}, {Ristorcelli}, {Rocha}, {Rosset}, {Rossetti}, {Roudier},
  {Rouill{\'e} d'Orfeuil}, {Rowan-Robinson}, {Rubi{\~n}o-Mart{\'\i}n},
  {Rusholme}, {Said}, {Salvatelli}, {Salvati}, {Sandri}, {Santos},
  {Savelainen}, {Savini}, {Scott}, {Seiffert}, {Serra}, {Shellard}, {Spencer},
  {Spinelli}, {Stolyarov}, {Stompor}, {Sudiwala}, {Sunyaev}, {Sutton},
  {Suur-Uski}, {Sygnet}, {Tauber}, {Terenzi}, {Toffolatti}, {Tomasi},
  {Tristram}, {Trombetti}, {Tucci}, {Tuovinen}, {T{\"u}rler}, {Umana},
  {Valenziano}, {Valiviita}, {Van Tent}, {Vielva}, {Villa}, {Wade}, {Wandelt},
  {Wehus}, {White}, {White}, {Wilkinson}, {Yvon}, {Zacchei}, \&
  {Zonca}}]{Planck2016}
{Ade}, P.~A.~R., {Aghanim}, N., {Arnaud}, M., {et~al.} 2016, \aap, 594, A13,
  \dodoi{10.1051/0004-6361/201525830}

\bibitem[{{Ajith} {et~al.}(2011){Ajith}, {Hannam}, {Husa}, {Chen},
  {Br{\"u}gmann}, {Dorband}, {M{\"u}ller}, {Ohme}, {Pollney}, {Reisswig}, \&
  et~al.}]{Ajith2011}
{Ajith}, P., {Hannam}, M., {Husa}, S., {et~al.} 2011, \prl, 106, 241101,
  \dodoi{10.1103/PhysRevLett.106.241101}

\bibitem[{{Antoni} \& {Quataert}(2022)}]{Antoni2022}
{Antoni}, A., \& {Quataert}, E. 2022, \mnras, 511, 176,
  \dodoi{10.1093/mnras/stab3776}

\bibitem[{{Asplund} {et~al.}(2009){Asplund}, {Grevesse}, {Sauval}, \&
  {Scott}}]{2009Asplund}
{Asplund}, M., {Grevesse}, N., {Sauval}, A.~J., \& {Scott}, P. 2009, \araa, 47,
  481, \dodoi{10.1146/annurev.astro.46.060407.145222}

\bibitem[{{Batta} \& {Ramirez-Ruiz}(2019)}]{Batta2019}
{Batta}, A., \& {Ramirez-Ruiz}, E. 2019, arXiv e-prints, arXiv:1904.04835.
\newblock \doarXiv{1904.04835}

\bibitem[{{Batta} {et~al.}(2017){Batta}, {Ramirez-Ruiz}, \&
  {Fryer}}]{Batta2017}
{Batta}, A., {Ramirez-Ruiz}, E., \& {Fryer}, C. 2017, \apjl, 846, L15,
  \dodoi{10.3847/2041-8213/aa8506}

\bibitem[{{Bavera} {et~al.}(2020){Bavera}, {Fragos}, {Qin}, {Zapartas},
  {Neijssel}, {Mandel}, {Batta}, {Gaebel}, {Kimball}, \&
  {Stevenson}}]{Bavera2020_origin_spin_BBHs}
{Bavera}, S.~S., {Fragos}, T., {Qin}, Y., {et~al.} 2020, \aap, 635, A97,
  \dodoi{10.1051/0004-6361/201936204}

\bibitem[{Bavera {et~al.}(2021)Bavera, Fragos, Zevin, Berry, Marchant, Andrews,
  Coughlin, Dotter, Kovlakas, Misra, Serra-Perez, Qin, Rocha, RomÃ¡n-Garza,
  Tran, \& Zapartas}]{bavera2021_impact_of_MT}
Bavera, S.~S., Fragos, T., Zevin, M., {et~al.} 2021, A\&A,
  \dodoi{10.1051/0004-6361/202039804}

\bibitem[{{Begelman}(2002)}]{Begelman2002}
{Begelman}, M.~C. 2002, \apjl, 568, L97, \dodoi{10.1086/340457}

\bibitem[{{Belczynski} {et~al.}(2011){Belczynski}, {Bulik}, \&
  {Bailyn}}]{Belczynski2011}
{Belczynski}, K., {Bulik}, T., \& {Bailyn}, C. 2011, \apjl, 742, L2,
  \dodoi{10.1088/2041-8205/742/1/L2}

\bibitem[{{Belczynski} {et~al.}(2012){Belczynski}, {Bulik}, \&
  {Fryer}}]{Belczynski2012}
{Belczynski}, K., {Bulik}, T., \& {Fryer}, C.~L. 2012, arXiv e-prints,
  arXiv:1208.2422.
\newblock \doarXiv{1208.2422}

\bibitem[{{Belczynski} {et~al.}(2008){Belczynski}, {Kalogera}, {Rasio}, {Taam},
  {Zezas}, {Bulik}, {Maccarone}, \& {Ivanova}}]{belczynski_compact_2008}
{Belczynski}, K., {Kalogera}, V., {Rasio}, F.~A., {et~al.} 2008, \apjs, 174,
  223, \dodoi{10.1086/521026}

\bibitem[{{Breivik} {et~al.}(2020){Breivik}, {Coughlin}, {Zevin}, {Rodriguez},
  {Kremer}, {Ye}, {Andrews}, {Kurkowski}, {Digman}, {Larson}, \&
  {Rasio}}]{Breivik2020}
{Breivik}, K., {Coughlin}, S., {Zevin}, M., {et~al.} 2020, \apj, 898, 71,
  \dodoi{10.3847/1538-4357/ab9d85}

\bibitem[{{Briel} {et~al.}(2022){Briel}, {Stevance}, \& {Eldridge}}]{Briel2022}
{Briel}, M.~M., {Stevance}, H.~F., \& {Eldridge}, J.~J. 2022, arXiv e-prints,
  arXiv:2206.13842.
\newblock \doarXiv{2206.13842}

\bibitem[{{Callister} {et~al.}(2022){Callister}, {Miller}, {Chatziioannou}, \&
  {Farr}}]{Callister2022}
{Callister}, T.~A., {Miller}, S.~J., {Chatziioannou}, K., \& {Farr}, W.~M.
  2022, arXiv e-prints, arXiv:2205.08574.
\newblock \doarXiv{2205.08574}

\bibitem[{{Cantiello} {et~al.}(2014){Cantiello}, {Mankovich}, {Bildsten},
  {Christensen-Dalsgaard}, \& {Paxton}}]{Cantiello2014}
{Cantiello}, M., {Mankovich}, C., {Bildsten}, L., {Christensen-Dalsgaard}, J.,
  \& {Paxton}, B. 2014, \apj, 788, 93, \dodoi{10.1088/0004-637X/788/1/93}

\bibitem[{{Claeys} {et~al.}(2014){Claeys}, {Pols}, {Izzard}, {Vink}, \&
  {Verbunt}}]{Claeys2014}
{Claeys}, J.~S.~W., {Pols}, O.~R., {Izzard}, R.~G., {Vink}, J., \& {Verbunt},
  F.~W.~M. 2014, \aap, 563, A83, \dodoi{10.1051/0004-6361/201322714}

\bibitem[{{de Mink} \& {Belczynski}(2015)}]{deMink2015}
{de Mink}, S.~E., \& {Belczynski}, K. 2015, \apj, 814, 58,
  \dodoi{10.1088/0004-637X/814/1/58}

\bibitem[{{de Mink} \& {Mandel}(2016)}]{deMink2016}
{de Mink}, S.~E., \& {Mandel}, I. 2016, \mnras, 460, 3545,
  \dodoi{10.1093/mnras/stw1219}

\bibitem[{{Eldridge} {et~al.}(2017){Eldridge}, {Stanway}, {Xiao}, {McClelland},
  {Taylor}, {Ng}, {Greis}, \& {Bray}}]{Eldridge2017}
{Eldridge}, J.~J., {Stanway}, E.~R., {Xiao}, L., {et~al.} 2017, \pasa, 34,
  e058, \dodoi{10.1017/pasa.2017.51}

\bibitem[{{Ertl} {et~al.}(2020){Ertl}, {Woosley}, {Sukhbold}, \&
  {Janka}}]{Ertl2020}
{Ertl}, T., {Woosley}, S.~E., {Sukhbold}, T., \& {Janka}, H.~T. 2020, \apj,
  890, 51, \dodoi{10.3847/1538-4357/ab6458}

\bibitem[{{Fishbach} \& {Kalogera}(2022)}]{Fishbach2022}
{Fishbach}, M., \& {Kalogera}, V. 2022, \apjl, 929, L26,
  \dodoi{10.3847/2041-8213/ac64a5}

\bibitem[{{Fragos} \& {McClintock}(2015)}]{FragosMcClintock2015}
{Fragos}, T., \& {McClintock}, J.~E. 2015, \apj, 800, 17,
  \dodoi{10.1088/0004-637X/800/1/17}

\bibitem[{{Fragos} {et~al.}(2022){Fragos}, {Andrews}, {Bavera}, {Berry},
  {Coughlin}, {Dotter}, {Giri}, {Kalogera}, {Katsaggelos}, {Kovlakas},
  {Lalvani}, {Misra}, {Srivastava}, {Qin}, {Rocha}, {Roman-Garza}, {Serra},
  {Stahle}, {Sun}, {Teng}, {Trajcevski}, {Hai Tran}, {Xing}, {Zapartas}, \&
  {Zevin}}]{Fragos2022}
{Fragos}, T., {Andrews}, J.~J., {Bavera}, S.~S., {et~al.} 2022, arXiv e-prints,
  arXiv:2202.05892.
\newblock \doarXiv{2202.05892}

\bibitem[{{Fryer} {et~al.}(2012){Fryer}, {Belczynski}, {Wiktorowicz},
  {Dominik}, {Kalogera}, \& {Holz}}]{Fryer2012}
{Fryer}, C.~L., {Belczynski}, K., {Wiktorowicz}, G., {et~al.} 2012, \apj, 749,
  91, \dodoi{10.1088/0004-637X/749/1/91}

\bibitem[{{Fuller} {et~al.}(2015){Fuller}, {Cantiello}, {Lecoanet}, \&
  {Quataert}}]{Fuller2015}
{Fuller}, J., {Cantiello}, M., {Lecoanet}, D., \& {Quataert}, E. 2015, \apj,
  810, 101, \dodoi{10.1088/0004-637X/810/2/101}

\bibitem[{{Fuller} \& {Lu}(2022)}]{Fuller2022}
{Fuller}, J., \& {Lu}, W. 2022, \mnras, 511, 3951,
  \dodoi{10.1093/mnras/stac317}

\bibitem[{{Fuller} \& {Ma}(2019)}]{FullerMa2019}
{Fuller}, J., \& {Ma}, L. 2019, \apjl, 881, L1,
  \dodoi{10.3847/2041-8213/ab339b}

\bibitem[{{Fuller} {et~al.}(2019){Fuller}, {Piro}, \&
  {Jermyn}}]{Fuller2019_slowing_spins}
{Fuller}, J., {Piro}, A.~L., \& {Jermyn}, A.~S. 2019, \mnras, 485, 3661,
  \dodoi{10.1093/mnras/stz514}

\bibitem[{{Galaudage} {et~al.}(2021){Galaudage}, {Talbot}, {Nagar}, {Jain},
  {Thrane}, \& {Mandel}}]{Galaudage2021}
{Galaudage}, S., {Talbot}, C., {Nagar}, T., {et~al.} 2021, \apjl, 921, L15,
  \dodoi{10.3847/2041-8213/ac2f3c}

\bibitem[{{Gallegos-Garcia} {et~al.}(2021){Gallegos-Garcia}, {Berry},
  {Marchant}, \& {Kalogera}}]{Gallegos-Garcia2021}
{Gallegos-Garcia}, M., {Berry}, C. P.~L., {Marchant}, P., \& {Kalogera}, V.
  2021, \apj, 922, 110, \dodoi{10.3847/1538-4357/ac2610}

\bibitem[{{Hamann} \& {Koesterke}(1998)}]{Hamann1998}
{Hamann}, W.~R., \& {Koesterke}, L. 1998, \aap, 335, 1003

\bibitem[{{Heger} {et~al.}(2005){Heger}, {Woosley}, \& {Spruit}}]{Heger2005}
{Heger}, A., {Woosley}, S.~E., \& {Spruit}, H.~C. 2005, \apj, 626, 350,
  \dodoi{10.1086/429868}

\bibitem[{{Hirai} \& {Mandel}(2021)}]{Hirai2021}
{Hirai}, R., \& {Mandel}, I. 2021, \pasa, 38, e056,
  \dodoi{10.1017/pasa.2021.53}

\bibitem[{{Hunter}(2007)}]{Hunter2007}
{Hunter}, J.~D. 2007, Computing in Science and Engineering, 9, 90,
  \dodoi{10.1109/MCSE.2007.55}

\bibitem[{{Hurley} {et~al.}(2000){Hurley}, {Pols}, \& {Tout}}]{Hurley2000}
{Hurley}, J.~R., {Pols}, O.~R., \& {Tout}, C.~A. 2000, \mnras, 315, 543,
  \dodoi{10.1046/j.1365-8711.2000.03426.x}

\bibitem[{{Hurley} {et~al.}(2002){Hurley}, {Tout}, \& {Pols}}]{Hurley2002}
{Hurley}, J.~R., {Tout}, C.~A., \& {Pols}, O.~R. 2002, \mnras, 329, 897,
  \dodoi{10.1046/j.1365-8711.2002.05038.x}

\bibitem[{{Ivanova} {et~al.}(2013){Ivanova}, {Justham}, {Chen}, {De Marco},
  {Fryer}, {Gaburov}, {Ge}, {Glebbeek}, {Han}, {Li}, \& et~al.}]{Ivanova2013}
{Ivanova}, N., {Justham}, S., {Chen}, X., {et~al.} 2013, \aapr, 21, 59,
  \dodoi{10.1007/s00159-013-0059-2}

\bibitem[{{Kimball} {et~al.}(2020){Kimball}, {Talbot}, {Berry}, {Carney},
  {Zevin}, {Thrane}, \& {Kalogera}}]{Kimball2020}
{Kimball}, C., {Talbot}, C., {Berry}, C. P.~L., {et~al.} 2020, \apj, 900, 177,
  \dodoi{10.3847/1538-4357/aba518}

\bibitem[{{Kimball} {et~al.}(2021){Kimball}, {Talbot}, {Berry}, {Zevin},
  {Thrane}, {Kalogera}, {Buscicchio}, {Carney}, {Dent}, {Middleton}, \&
  et~al.}]{Kimball2021}
---. 2021, \apjl, 915, L35, \dodoi{10.3847/2041-8213/ac0aef}

\bibitem[{{King} \& {Kolb}(1999)}]{King1999a}
{King}, A.~R., \& {Kolb}, U. 1999, \mnras, 305, 654,
  \dodoi{10.1046/j.1365-8711.1999.02482.x}

\bibitem[{{Klencki} {et~al.}(2018){Klencki}, {Moe}, {Gladysz}, {Chruslinska},
  {Holz}, \& {Belczynski}}]{KlenckiMoe2018}
{Klencki}, J., {Moe}, M., {Gladysz}, W., {et~al.} 2018, \aap, 619, A77,
  \dodoi{10.1051/0004-6361/201833025}

\bibitem[{{Liotine} {et~al.}(2022){Liotine}, {Zevin}, {Berry}, {Doctor}, \&
  {Kalogera}}]{Liotine2022}
{Liotine}, C., {Zevin}, M., {Berry}, C., {Doctor}, Z., \& {Kalogera}, V. 2022,
  arXiv e-prints, arXiv:2210.01825.
\newblock \doarXiv{2210.01825}

\bibitem[{{Liu} {et~al.}(2008){Liu}, {McClintock}, {Narayan}, {Davis}, \&
  {Orosz}}]{Liu2008}
{Liu}, J., {McClintock}, J.~E., {Narayan}, R., {Davis}, S.~W., \& {Orosz},
  J.~A. 2008, \apjl, 679, L37, \dodoi{10.1086/588840}

\bibitem[{{Mandel} \& {de Mink}(2016)}]{Mandel2016}
{Mandel}, I., \& {de Mink}, S.~E. 2016, \mnras, 458, 2634,
  \dodoi{10.1093/mnras/stw379}

\bibitem[{{Mandel} \& {Fragos}(2020)}]{Mandel2020a}
{Mandel}, I., \& {Fragos}, T. 2020, \apjl, 895, L28,
  \dodoi{10.3847/2041-8213/ab8e41}

\bibitem[{{Marchant} {et~al.}(2016){Marchant}, {Langer}, {Podsiadlowski},
  {Tauris}, \& {Moriya}}]{Marchant2016}
{Marchant}, P., {Langer}, N., {Podsiadlowski}, P., {Tauris}, T.~M., \&
  {Moriya}, T.~J. 2016, \aap, 588, A50, \dodoi{10.1051/0004-6361/201628133}

\bibitem[{{M}c{K}inney(2010)}]{mckinney-proc-scipy-2010}
{M}c{K}inney. 2010, in {P}roceedings of the 9th {P}ython in {S}cience
  {C}onference, ed. {S}t\'efan van~der {W}alt \& {J}arrod {M}illman, 56 -- 61,
  \dodoi{10.25080/Majora-92bf1922-00a}

\bibitem[{{McKinney} {et~al.}(2014){McKinney}, {Tchekhovskoy}, {Sadowski}, \&
  {Narayan}}]{McKinney2014}
{McKinney}, J.~C., {Tchekhovskoy}, A., {Sadowski}, A., \& {Narayan}, R. 2014,
  \mnras, 441, 3177, \dodoi{10.1093/mnras/stu762}

\bibitem[{{Miller-Jones} {et~al.}(2021){Miller-Jones}, {Bahramian}, {Orosz},
  {Mandel}, {Gou}, {Maccarone}, {Neijssel}, {Zhao}, {Zi{\'o}{\l}kowski},
  {Reid}, {Uttley}, {Zheng}, {Byun}, {Dodson}, {Grinberg}, {Jung}, {Kim},
  {Marcote}, {Markoff}, {Rioja}, {Rushton}, {Russell}, {Sivakoff}, {Tetarenko},
  {Tudose}, \& {Wilms}}]{Miller-Jones2021}
{Miller-Jones}, J. C.~A., {Bahramian}, A., {Orosz}, J.~A., {et~al.} 2021,
  Science, 371, 1046, \dodoi{10.1126/science.abb3363}

\bibitem[{{Moe} \& {Di Stefano}(2017)}]{MoeDiStefano2017}
{Moe}, M., \& {Di Stefano}, R. 2017, \apjs, 230, 15,
  \dodoi{10.3847/1538-4365/aa6fb6}

\bibitem[{{Mould} {et~al.}(2022){Mould}, {Gerosa}, {Broekgaarden}, \&
  {Steinle}}]{Mould2022}
{Mould}, M., {Gerosa}, D., {Broekgaarden}, F.~S., \& {Steinle}, N. 2022, arXiv
  e-prints, arXiv:2205.12329.
\newblock \doarXiv{2205.12329}

\bibitem[{{Neijssel} {et~al.}(2021){Neijssel}, {Vinciguerra},
  {Vigna-G{\'o}mez}, {Hirai}, {Miller-Jones}, {Bahramian}, {Maccarone}, \&
  {Mandel}}]{Neijssel2021_cygX1}
{Neijssel}, C.~J., {Vinciguerra}, S., {Vigna-G{\'o}mez}, A., {et~al.} 2021,
  \apj, 908, 118, \dodoi{10.3847/1538-4357/abde4a}

\bibitem[{{Neijssel} {et~al.}(2019){Neijssel}, {Vigna-G{\'o}mez}, {Stevenson},
  {Barrett}, {Gaebel}, {Broekgaarden}, {de Mink}, {Sz{\'e}csi}, {Vinciguerra},
  \& {Mandel}}]{neijssel_effect_2019}
{Neijssel}, C.~J., {Vigna-G{\'o}mez}, A., {Stevenson}, S., {et~al.} 2019,
  \mnras, 490, 3740, \dodoi{10.1093/mnras/stz2840}

\bibitem[{{Olejak} {et~al.}(2021){Olejak}, {Belczynski}, \&
  {Ivanova}}]{olejak2021impact_CE_criteria}
{Olejak}, A., {Belczynski}, K., \& {Ivanova}, N. 2021, \aap, 651, A100,
  \dodoi{10.1051/0004-6361/202140520}

\bibitem[{{Pavlovskii} {et~al.}(2017){Pavlovskii}, {Ivanova}, {Belczynski}, \&
  {Van}}]{Pavlovskii2017}
{Pavlovskii}, K., {Ivanova}, N., {Belczynski}, K., \& {Van}, K.~X. 2017,
  \mnras, 465, 2092, \dodoi{10.1093/mnras/stw2786}

\bibitem[{{Paxton} {et~al.}(2011){Paxton}, {Bildsten}, {Dotter}, {Herwig},
  {Lesaffre}, \& {Timmes}}]{Paxton2011}
{Paxton}, B., {Bildsten}, L., {Dotter}, A., {et~al.} 2011, \apjs, 192, 3,
  \dodoi{10.1088/0067-0049/192/1/3}

\bibitem[{{Paxton} {et~al.}(2013){Paxton}, {Cantiello}, {Arras}, {Bildsten},
  {Brown}, {Dotter}, {Mankovich}, {Montgomery}, {Stello}, {Timmes}, \&
  {Townsend}}]{Paxton2013}
{Paxton}, B., {Cantiello}, M., {Arras}, P., {et~al.} 2013, \apjs, 208, 4,
  \dodoi{10.1088/0067-0049/208/1/4}

\bibitem[{{Paxton} {et~al.}(2015){Paxton}, {Marchant}, {Schwab}, {Bauer},
  {Bildsten}, {Cantiello}, {Dessart}, {Farmer}, {Hu}, {Langer}, {Townsend},
  {Townsley}, \& {Timmes}}]{Paxton2015}
{Paxton}, B., {Marchant}, P., {Schwab}, J., {et~al.} 2015, \apjs, 220, 15,
  \dodoi{10.1088/0067-0049/220/1/15}

\bibitem[{{Paxton} {et~al.}(2019){Paxton}, {Smolec}, {Schwab}, {Gautschy},
  {Bildsten}, {Cantiello}, {Dotter}, {Farmer}, {Goldberg}, {Jermyn}, {Kanbur},
  {Marchant}, {Thoul}, {Townsend}, {Wolf}, {Zhang}, \& {Timmes}}]{Paxton2019}
{Paxton}, B., {Smolec}, R., {Schwab}, J., {et~al.} 2019, \apjs, 243, 10,
  \dodoi{10.3847/1538-4365/ab2241}

\bibitem[{{Podsiadlowski} {et~al.}(2003){Podsiadlowski}, {Rappaport}, \&
  {Han}}]{Podsiadlowski2003}
{Podsiadlowski}, P., {Rappaport}, S., \& {Han}, Z. 2003, \mnras, 341, 385,
  \dodoi{10.1046/j.1365-8711.2003.06464.x}

\bibitem[{{Qin} {et~al.}(2018){Qin}, {Fragos}, {Meynet}, {Andrews},
  {S{\o}rensen}, \& {Song}}]{Qin2018}
{Qin}, Y., {Fragos}, T., {Meynet}, G., {et~al.} 2018, \aap, 616, A28,
  \dodoi{10.1051/0004-6361/201832839}

\bibitem[{{Qin} {et~al.}(2019){Qin}, {Marchant}, {Fragos}, {Meynet}, \&
  {Kalogera}}]{Qin2019}
{Qin}, Y., {Marchant}, P., {Fragos}, T., {Meynet}, G., \& {Kalogera}, V. 2019,
  \apjl, 870, L18, \dodoi{10.3847/2041-8213/aaf97b}

\bibitem[{{Qin} {et~al.}(2022){Qin}, {Shu}, {Yi}, \& {Wang}}]{Qin2022}
{Qin}, Y., {Shu}, X., {Yi}, S., \& {Wang}, Y.-Z. 2022, Research in Astronomy
  and Astrophysics, 22, 035023, \dodoi{10.1088/1674-4527/ac4ca4}

\bibitem[{{Remillard} \& {McClintock}(2006)}]{Remillard2006}
{Remillard}, R.~A., \& {McClintock}, J.~E. 2006, \araa, 44, 49,
  \dodoi{10.1146/annurev.astro.44.051905.092532}

\bibitem[{{Reynolds}(2021)}]{Reynolds2021}
{Reynolds}, C.~S. 2021, \araa, 59, \dodoi{10.1146/annurev-astro-112420-035022}

\bibitem[{{Roulet} {et~al.}(2021){Roulet}, {Chia}, {Olsen}, {Dai},
  {Venumadhav}, {Zackay}, \& {Zaldarriaga}}]{Roulet2021}
{Roulet}, J., {Chia}, H.~S., {Olsen}, S., {et~al.} 2021, \prd, 104, 083010,
  \dodoi{10.1103/PhysRevD.104.083010}

\bibitem[{{Roulet} \& {Zaldarriaga}(2019)}]{RouletZaldarriaga2019}
{Roulet}, J., \& {Zaldarriaga}, M. 2019, \mnras, 484, 4216,
  \dodoi{10.1093/mnras/stz226}

\bibitem[{{Santamar{\'\i}a} {et~al.}(2010){Santamar{\'\i}a}, {Ohme}, {Ajith},
  {Br{\"u}gmann}, {Dorband}, {Hannam}, {Husa}, {M{\"o}sta}, {Pollney},
  {Reisswig}, \& et~al.}]{Santamaria2010}
{Santamar{\'\i}a}, L., {Ohme}, F., {Ajith}, P., {et~al.} 2010, \prd, 82,
  064016, \dodoi{10.1103/PhysRevD.82.064016}

\bibitem[{{Schr{\o}der} {et~al.}(2018){Schr{\o}der}, {Batta}, \&
  {Ramirez-Ruiz}}]{Schroder2018}
{Schr{\o}der}, S.~L., {Batta}, A., \& {Ramirez-Ruiz}, E. 2018, \apjl, 862, L3,
  \dodoi{10.3847/2041-8213/aacf8d}

\bibitem[{{Shao} \& {Li}(2022)}]{Shao2022}
{Shao}, Y., \& {Li}, X.-D. 2022, \apj, 930, 26,
  \dodoi{10.3847/1538-4357/ac61da}

\bibitem[{{Song} {et~al.}(2016){Song}, {Meynet}, {Maeder}, {Ekstr{\"o}m}, \&
  {Eggenberger}}]{Song2016}
{Song}, H.~F., {Meynet}, G., {Maeder}, A., {Ekstr{\"o}m}, S., \& {Eggenberger},
  P. 2016, \aap, 585, A120, \dodoi{10.1051/0004-6361/201526074}

\bibitem[{{Spruit}(1999)}]{Spruit1999}
{Spruit}, H.~C. 1999, \aap, 349, 189.
\newblock \doarXiv{astro-ph/9907138}

\bibitem[{{Stanway} \& {Eldridge}(2018)}]{StanwayEldridge2018}
{Stanway}, E.~R., \& {Eldridge}, J.~J. 2018, \mnras, 479, 75,
  \dodoi{10.1093/mnras/sty1353}

\bibitem[{{Suijs} {et~al.}(2008){Suijs}, {Langer}, {Poelarends}, {Yoon},
  {Heger}, \& {Herwig}}]{Suijs2008}
{Suijs}, M.~P.~L., {Langer}, N., {Poelarends}, A.~J., {et~al.} 2008, \aap, 481,
  L87, \dodoi{10.1051/0004-6361:200809411}

\bibitem[{{Sukhbold} {et~al.}(2016){Sukhbold}, {Ertl}, {Woosley}, {Brown}, \&
  {Janka}}]{Sukhbold2016}
{Sukhbold}, T., {Ertl}, T., {Woosley}, S.~E., {Brown}, J.~M., \& {Janka}, H.~T.
  2016, \apj, 821, 38, \dodoi{10.3847/0004-637X/821/1/38}

\bibitem[{{Tauris} {et~al.}(2000){Tauris}, {van den Heuvel}, \&
  {Savonije}}]{Tauris2000}
{Tauris}, T.~M., {van den Heuvel}, E. P.~J., \& {Savonije}, G.~J. 2000, \apjl,
  530, L93, \dodoi{10.1086/312496}

\bibitem[{{Valsecchi} {et~al.}(2010){Valsecchi}, {Glebbeek}, {Farr}, {Fragos},
  {Willems}, {Orosz}, {Liu}, \& {Kalogera}}]{Valsecchi2010Nature}
{Valsecchi}, F., {Glebbeek}, E., {Farr}, W.~M., {et~al.} 2010, \nat, 468, 77,
  \dodoi{10.1038/nature09463}

\bibitem[{{van den Heuvel}(1976)}]{vandenHeuvel1976}
{van den Heuvel}, E.~P.~J. 1976, in IAU Symposium, Vol.~73, Structure and
  Evolution of Close Binary Systems, ed. P.~{Eggleton}, S.~{Mitton}, \&
  J.~{Whelan}, 35

\bibitem[{{van den Heuvel}(2019)}]{vandenHeuvel2019}
{van den Heuvel}, E. P.~J. 2019, IAU Symposium, 346, 1,
  \dodoi{10.1017/S1743921319001315}

\bibitem[{{van der Walt} {et~al.}(2011){van der Walt}, {Colbert}, \&
  {Varoquaux}}]{vanderwalt2011}
{van der Walt}, S., {Colbert}, S.~C., \& {Varoquaux}, G. 2011, Computing in
  Science and Engineering, 13, 22, \dodoi{10.1109/MCSE.2011.37}

\bibitem[{{van Son} {et~al.}(2020){van Son}, {De Mink}, {Broekgaarden},
  {Renzo}, {Justham}, {Laplace}, {Mor{\'a}n-Fraile}, {Hendriks}, \&
  {Farmer}}]{vanSon2020}
{van Son}, L.~A.~C., {De Mink}, S.~E., {Broekgaarden}, F.~S., {et~al.} 2020,
  \apj, 897, 100, \dodoi{10.3847/1538-4357/ab9809}

\bibitem[{{Vink} \& {de Koter}(2005)}]{Vink2005}
{Vink}, J.~S., \& {de Koter}, A. 2005, \aap, 442, 587,
  \dodoi{10.1051/0004-6361:20052862}

\bibitem[{{Vink} {et~al.}(2001){Vink}, {de Koter}, \&
  {Lamers}}]{Vink2001_hot_highH}
{Vink}, J.~S., {de Koter}, A., \& {Lamers}, H.~J.~G.~L.~M. 2001, \aap, 369,
  574, \dodoi{10.1051/0004-6361:20010127}

\bibitem[{{Webbink}(1984)}]{Webbink1984}
{Webbink}, R.~F. 1984, \apj, 277, 355, \dodoi{10.1086/161701}

\bibitem[{{Yoon} {et~al.}(2010){Yoon}, {Woosley}, \& {Langer}}]{Yoon2010}
{Yoon}, S.~C., {Woosley}, S.~E., \& {Langer}, N. 2010, \apj, 725, 940,
  \dodoi{10.1088/0004-637X/725/1/940}

\bibitem[{{Zevin} \& {Bavera}(2022)}]{ZevinBavera2022}
{Zevin}, M., \& {Bavera}, S.~S. 2022, \apj, 933, 86,
  \dodoi{10.3847/1538-4357/ac6f5d}

\end{thebibliography}
\bibliographystyle{aasjournal}

\end{document}